%% file: main.tex
\documentclass{SciPost}

\hypersetup{
    colorlinks,
    linkcolor={red!50!black},
    citecolor={blue!50!black},
    urlcolor={blue!80!black},
	pdftitle={Reentrances Phase Diagram - draft},
	pdfauthor={LG PV CC}
	bookmarks=true,
	bookmarksopen=true,
}

\usepackage[bitstream-charter]{mathdesign}
\graphicspath{{./Figures/}} 
\usepackage[italian, english]{babel} 
\usepackage[T1]{fontenc}
\usepackage{tikz}
\usetikzlibrary{shapes.arrows}
\usepackage{adjustbox}
\usetikzlibrary {quotes}
\tikzset{%
  highlighta/.style={rectangle,rounded corners,fill=red!15,draw,
    fill opacity=0.5,inner sep=0pt}
}
\usepackage{tikz-network}
\usetikzlibrary{positioning}
\tikzset{%
  highlightb/.style={rectangle,rounded corners,fill=blue!15,draw,
    fill opacity=0.5,inner sep=0pt}
}
\tikzset{%
  highlightc/.style={rectangle,rounded corners,fill=green!15,draw,
    fill opacity=0.5,inner sep=0pt}
}\tikzset{%
  highlightd/.style={rectangle,rounded corners,fill=yellow!15,draw,
    fill opacity=0.5,inner sep=0pt}
}
\newcommand{\tikzmark}[2]{\tikz[overlay,remember picture,
  baseline=(#1.base)] \node (#1) {#2};}
\newcommand{\Highlighta}[3][submatrix]{%
    \tikz[overlay,remember picture]{
    \node[highlighta,fit=(#2.north west) (#3.south east)] (#1) {};}
}
\newcommand{\Highlightb}[3][submatrix]{%
    \tikz[overlay,remember picture]{
    \node[highlightb,fit=(#2.north west) (#3.south east)] (#1) {};}
}
\newcommand{\Highlightc}[3][submatrix]{%
    \tikz[overlay,remember picture]{
    \node[highlightc,fit=(#2.north west) (#3.south east)] (#1) {};}
}
\newcommand{\Highlightd}[3][submatrix]{%
    \tikz[overlay,remember picture]{
    \node[highlightd,fit=(#2.north west) (#3.south east)] (#1) {};}
}

\usepackage{subcaption}

\newcommand{\matrixbroc}[1]{\mathbf{#1}}

\newcommand{\order}[1]{\mathcal{O}\left(#1\right)}
\newcommand{\conj}[1]{\bar{#1}}

\newcommand{\wirtinger}{\partial_{\conj{z}}}
\newcommand{\signop}{{\rm sign}}

\newcommand{\figref}[1]{Fig.~\ref{#1}}
\newcommand{\ie}{\emph{i.e.}}
\newcommand{\multilinecomment}[1]{} 
\newcommand{\ER}{Erd\"os-R\'enyi }
\newcommand{\id}[1]{\textbf{1}_{#1}}
\newcommand{\Tr}{\mbox{Tr}}
\renewcommand{\i}{\mathrm{i}}
\newcommand{\ub}[1]{\underline{\textbf{#1}}}
\newcommand{\G}{\underline{G}}
\newcommand{\M}{\underline{M}}
\newcommand{\z}{\underline{z}_{\eta}}
\newcommand{\Ht}{\tilde{\underline{H}}}
\newcommand{\vs}{\mbox{vs.}}

\NewDocumentCommand{\dd}{o m}{\mathop{}\!\mathrm{d}\IfValueT{#1}{^#1\mkern-1mu}#2}
\newcommand{\der}[3][]{\ensuremath{\frac{\dd^{#1} #2}{\dd #3}}}
\newcommand{\pder}[3][]{\ensuremath{\frac{\partial^{#1} #2}{\partial #3}}}

\DeclareSymbolFont{usualmathcal}{OMS}{cmsy}{m}{n}
\DeclareSymbolFontAlphabet{\mathcal}{usualmathcal}

\fancypagestyle{SPstyle}{
\fancyhf{}
\lhead{\colorbox{scipostblue}{\bf \color{white} ~SciPost Physics }}
\rhead{{\bf \color{scipostdeepblue} ~Submission }}

\fancyfoot[C]{\textbf{\thepage}}
}

\begin{document}

\pagestyle{SPstyle}

\begin{center}{\Large \textbf{\color{scipostdeepblue}{
Spectral properties and phase diagrams of sparse antagonistic random matrices with diagonal disorder and Jacobian-like structure\\
}}}\end{center}

\begin{center}\textbf{
Luca Giammanco\textsuperscript{1$\star$},
Pietro Valigi\textsuperscript{2, 3$\dagger$} and
Chiara Cammarota\textsuperscript{2,4}
}\end{center}

\begin{center}
{\bf 1} Department of Computing Sciences, Bocconi University, 20136 Milano, Italy
\\
{\bf 2} Department of Physics, Sapienza University of Rome, P.le Aldo Moro 5, Rome, 00185, Italy
\\
{\bf 3} Department of Mathematical Sciences, University of Bath, Bath BA27AY, UK
\\
{\bf 4} INFN
\\[\baselineskip]
$\star$ \href{mailto:email1}{\small luca.giammanco@phd.unibocconi.it}\,,\quad
$\dagger$ \href{mailto:email2}{\small pv484@bath.ac.uk}
\end{center}

\section*{\color{scipostdeepblue}{Abstract}}
\textbf{\boldmath{
Complex interacting systems are often modelled by random matrices whose spectral properties dictate stability. In sparse antagonistic matrices without diagonal disorder, low connectivity gives rise to a characteristic reentrance effect in the spectral boundary near the real axis, which disappears via a continuous transition as the connectivity increases. The reentrance effect implies the presence of a complex leading eigenvalue, which suggests the existence of a phase characterized by oscillatory dynamics around equilibrium. Here, we expand the investigation to matrices featuring diagonal disorder and a Jacobian-like structure. In these settings, the spectrum also develops a segment of eigenvalues accumulating on the real axis, which can trigger a discontinuous jump of the complex leading eigenvalue to a purely real value. The interplay between connectivity and disorder produces a rich variety of spectral behaviours. Employing the cavity method and a an adaptation of the Population Dynamics algorithm, we map a phase diagram with five distinct spectral phases. Finally, we show that the algorithm underestimates the spectral support under strong disorder, motivating future technical developments to handle this limit.
}}


\vspace{\baselineskip}

\noindent\textcolor{white!90!black}{%
\fbox{\parbox{0.975\linewidth}{%
\textcolor{white!40!black}{\begin{tabular}{lr}%
  \begin{minipage}{0.6\textwidth}%
    {\small Copyright attribution to authors. \newline
    This work is a submission to SciPost Physics. \newline
    License information to appear upon publication. \newline
    Publication information to appear upon publication.}
  \end{minipage} & \begin{minipage}{0.4\textwidth}
    {\small Received Date \newline Accepted Date \newline Published Date}%
  \end{minipage}
\end{tabular}}
}}
}


\vspace{10pt}
\noindent\rule{\textwidth}{1pt}
\tableofcontents
\noindent\rule{\textwidth}{1pt}
\vspace{10pt}

\section{Introduction} \label{sec:introduction}

Complex interacting systems in physics, biology, and ecology can naturally be represented as networks, where nodes correspond to dynamical units and edges encode pairwise interactions. In this framework, graphs provide a flexible and unifying language to describe both the topology of interactions and their statistical properties. The spectral properties of matrices associated with such graphs provide direct insight into the dynamical behaviour of the corresponding models, including their stability and the nature of their response to perturbations~\cite{hasselblatt2003first, strogatz2024nonlinear}. For linear systems, this information is entirely encoded in the spectrum of the interaction matrix itself, whereas for more general non-linear models it is captured by the spectrum of the Jacobian matrix evaluated at a stationary state. In both cases, the real part of the leading eigenvalue, defined as the eigenvalue with the largest real part, determines stability: a negative real part implies linear stability, while a positive one signals an instability. On the other hand, if the leading eigenvalue has a nonzero imaginary part, then the principal mode of the system's response to a perturbation will be oscillatory. 

A paradigmatic example of how the properties of the interaction matrix can influence different kinds of stabilities is provided by the generalized Lotka–Volterra model \cite{bunin2017ecological, galla2018dynamically, biroli2018marginally}, presented in Appendix~\ref{app:GenLotka-Volterra}, which can be used to describe large interacting ecological communities. As shown in more detail in Appendix~\ref{app:stability_ecology}, the spectral properties of the interaction matrix encode information about different notions of stability, and even about the existence of fixed points~\cite{valigi2024local, stone2018feasibility, rohr2014structural}. In this ecological framework, different types of pairwise interactions between species are reflected in the sign structure of this asymmetric matrix. In particular, \emph{antagonistic} (or predator–prey) interactions, in which one species benefits at the expense of the other, correspond to sign-antisymmetric matrix elements. On the other hand, competitive or mutualistic interactions, which induce, respectively, a negative or positive effect on the growth of both species, are described by sign-symmetric matrix entries. Recent studies have shown that network topology, the sign structure of the interaction matrix, and disorder in the interactions significantly influence the spectral properties and, consequently, the system dynamics \cite{allesina2012stability, allesina2015predicting, poley2024eigenvalue, mambuca2022dynamical, valigi2024local}, both in dense and sparse models.

For dense random matrices, where each node interacts with a number of neighbours that grows with the system size $\order{N}$, classical dense random matrix theory provides a relatively complete picture. Under pretty general conditions the width of the spectrum in the complex plane in the large-$N$ limit grows as $\sqrt{N}$ and, for this reason, interactions are typically rescaled by $1/\sqrt{N}$ to ensure a bounded spectral support in the thermodynamic limit. More specifically, if the off-diagonal entries are independent and identically distributed with no correlation other than the one with the diagonally opposite element, the spectrum obeys the famous \emph{elliptical law} \cite{sommers1988spectrum, allesina2012stability} and the nature of the interactions only have a quantitative effect through the global correlation.

The situation changes qualitatively when interactions are sparse, namely when each node has a finite number of interactions whose average $c$ does not scale with $N$. In such case, rescaling interactions by $1/\sqrt{c}$ is not sufficient to suppress spectral growth, and stability properties depend sensitively on both network topology and interaction sign structure. Recent studies have shown that sparse random graphs with purely antagonistic interactions and a finite density of short loops (belonging to the broader class of so-called Locally Sign Stable graphs~\cite{valigi2024local}) exhibit a spectrum with finite support along the real axis, even in the thermodynamic limit~\cite{mambuca2022dynamical, valigi2024local}. As a result, stability is no longer critically affected by system size. By contrast, introducing a finite fraction of competitive or mutualistic interactions, or allowing for an extensive number of short loops, generates spectral tails that extend indefinitely along the real axis, restoring a size–stability trade-off~\cite{mambuca2022dynamical, valigi2024local, valigi2025spectraltails}.


At the same time, focusing on the bulk of the spectrum we can observe that, if the connectivity $c$ is large enough, its shape still resemble an ellipse, and it is also possible to compute the corrections from the elliptical shape through a perturbative expansion in terms of $1/c$ \cite{baron2025pathintegral}\footnote{Note however that the spectral tails are not captured by this perturbative approach~\cite{valigi2025spectraltails}.}. As we keep reducing $c$, the boundary deviates more and more from the elliptical shape~\cite{mambuca2022dynamical}. In particular, when the interactions are purely antagonistic and the connectivity is very small, the spectrum develops characteristic \emph{reentrances} close to the real axis~\cite{mambuca2022dynamical}: the spectral boundary changes concavity and shrinks, while the leading eigenvalue typically acquires a non-zero imaginary part. In linear systems, these spectral reentrances imply an oscillatory dynamics. As connectivity increases, these reentrances progressively reduce, leading to a continuous transition toward a phase in which the leading eigenvalue becomes real.

\begin{figure}[ht]
    \centering
    \includegraphics[width=0.85\textwidth]{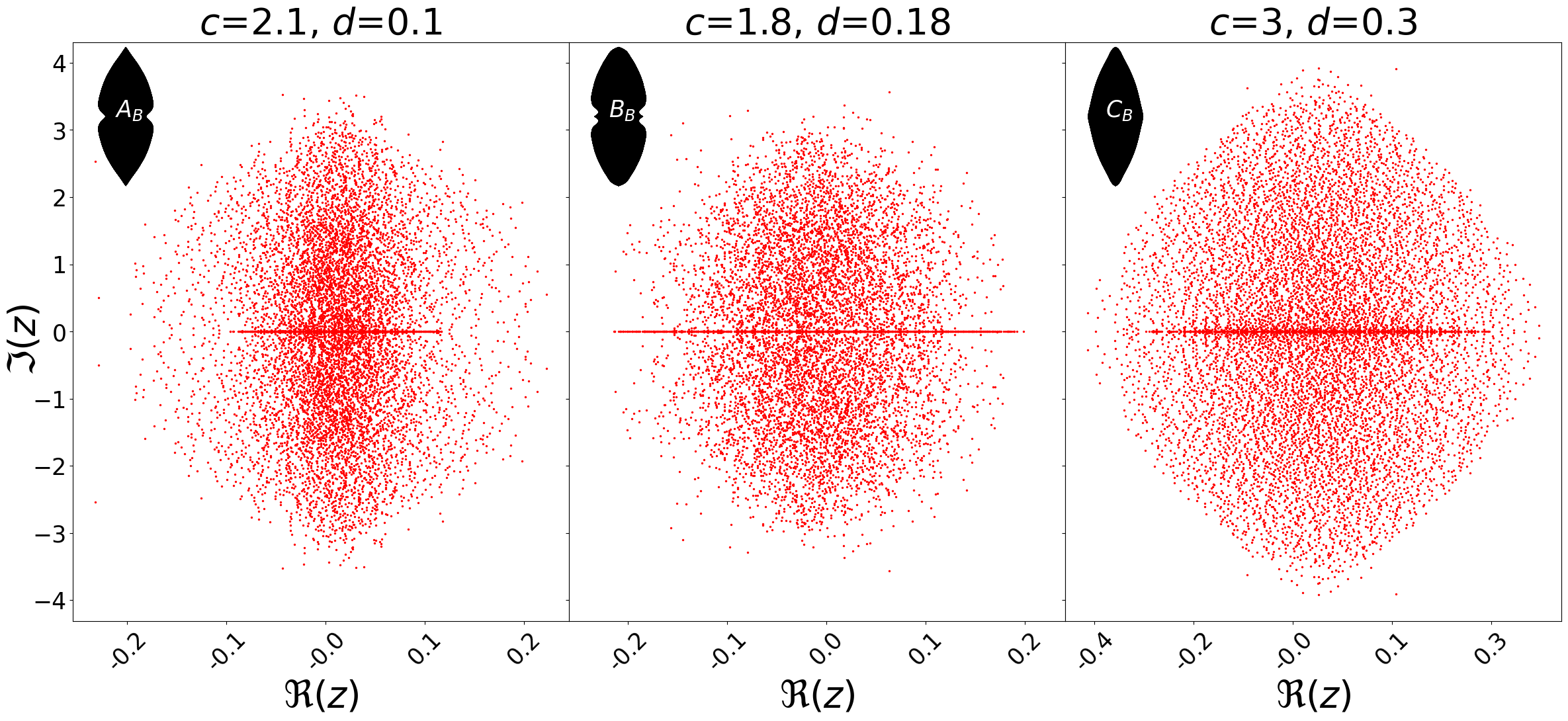}
    \caption{Examples of different shapes observed through direct diagonalisation of large sparse matrices. The matrices are built on an \ER adjacency matrix with size $N_D=10000$ and weights following the distribution described in Sec.~\ref{sec:interaction}; the values of the connectivity $c$ and diagonal disorder $d$ are reported in the titles of each picture. At low connectivity and diagonal disorder (left), we observe a concave spectral support with reentrances around the real axis and a line of concentrated eigenvalues contained \emph{inside the bulk}. Keeping the connectivity fixed and increasing the diagonal disorder (center), we observe that this segment starts to protrude from the bulk, affecting the shape of the spectral support. At larger connectivities (right) we generally observe convex spectral supports, and the segment is contained in the bulk of the spectrum unless the diagonal disorder is very strong. The black symbols located at the top-left of each plot are stylized figures that resemble the shape of the spectra: they must not be intended as rigorous representations, but rather as an aid to the reader for comparing the different figures in this paper.}
    \label{fig:initial_examples}
\end{figure}

More recently, it has been shown that these spectral reentrances persist even in the presence of diagonal disorder or Jacobian-like structure, which is characterised by correlations between elements belonging to the same row~\cite{valigi2024local}, thus potentially preserving the expectation of an associated oscillatory dynamics in proximity of the corresponding fixed points. In these cases, however, the spectrum also develops a segment of eigenvalues accumulating on the real axis. As the strength of diagonal disorder is increased, this segment elongates and a discontinuous transition may occur, at which the imaginary part of the leading eigenvalue jumps to zero. The interplay between connectivity-driven and disorder-driven transitions gives rise to a rich phenomenology, including multiple spectral phases characterized by qualitatively distinct shapes and stability properties. An example of three such phases can be observed in Fig.~\ref{fig:initial_examples}.

Despite this progress, a systematic characterization of these phases is hindered by the inherent difficulty of identifying the boundary in finite-size spectra obtained via direct diagonalization. To overcome this limitation, it is necessary to rely on semi-analytical approaches tailored to sparse asymmetric matrices, namely the \emph{cavity equations for the hermitised resolvent} \cite{metz2019spectral, mambuca2022dynamical, rogers2008cavity}. In this work, we solve such equations through a generalised \emph{Population Dynamics} algorithm \cite{mambuca2022dynamical}, extending existing methods to account for diagonal disorder and Jacobian-like correlations. This approach allows us to estimate the boundary of the spectrum in the thermodynamic limit and to construct a detailed phase diagram in the space of connectivity and disorder, identifying five different spectral phases.

The structure of the paper is as follows: in Section~\ref{sec:Model} we introduce the random matrix models we are going to investigate, introducing two different types of ensembles, namely Interaction-like (Sec.~\ref{sec:interaction}) and Jacobian-like matrices (Sec.~\ref{sec:jacobian}). In Section~\ref{sec:Algo} we discuss the method we use to estimate the boundary of the spectral distribution. In particular, in Sec.~\ref{sec:cavity_equations} we introduce the concept of trivial solution to the cavity equations for the hermitised resolvent and we explain how its linear stability can be investigated to estimate the spectral boundary. Then in Sec.~\ref{sec:PopDyn_Algo} we describe the Population Dynamics algorithm we employ to assess such stability. In Section~\ref{sec:Results}, we present the primary results of this work. Using the Population Dynamics algorithm, we map the spectral properties of the leading eigenvalues as a function of connectivity $c$ and diagonal disorder $d$. Specifically, we delineate and characterise five distinct phases, separated by various continuous and discontinuous transitions. Finally, in Section~\ref{sec:discussion} we discuss our findings and conclude. 

\section{Random Matrix models}\label{sec:Model}

Similarly to what has been done in Ref.~\cite{mambuca2022dynamical}, we will study random matrix ensembles inspired by the Generalized Lotka-Volterra equations in theoretical ecology. In particular, we will deal with random matrices $\textbf{A}$ of size $N\times N$ and of the form
\begin{equation}
    [\textbf{A}]_{ij}=\alpha_{ij}[\textbf{C}]_{ij}\,,
\end{equation}
where the $\{\alpha_{ij}\}$ are called weights or coupling strengths, while $\textbf{C}$ is the adjacency matrix of an \ER random graph~\cite{erdos1960evolution}, usually called the interaction network. An important parameter in our studies will be the connectivity $c$ of this graph, defined as the mean degree of the nodes:
\begin{equation}
c=\sum_{k=0}^{+\infty}p_{deg}(k)k\,,
\end{equation}
where $p_{deg}$ is the degree distribution of the network, corresponding to a Poisson distribution in the \ER case. As the name suggests, $c$ determines how "connected" the graph is, and it can be shown that at $c=1$ there is a discontinuous transition similar to a percolation one: below this value, the graph is typically divided in many small connected components, each containing a vanishing fraction of nodes, while at $c>1$ there also exists a "giant" connected component to which a nonzero fraction of nodes belongs \cite{erdos1960evolution}. 
We stress that even after this transition a nonvanishing fraction of finite disconnected components survives, including an extensive number of isolated nodes, which produce a trivial contribution to the spectrum. In the rest of the paper, we will ignore these isolated nodes, removing them from the matrices.

Having fixed the structure of the graph, the way we define the coupling strengths entirely determines the random matrix ensemble. We will focus on two cases, which can be referred to as \emph{Interaction-like} and \emph{Jacobian-like}. The form of these matrices is inspired by the generalized Lotka–Volterra model, although their relevance may extend well beyond this specific setting. For the interested reader, an introduction to such ecological model and to the two matrices governing its different types of stability is provided in Appendix~\ref{app:GenLotka-Volterra}.

\subsection{Interaction-like matrices}\label{sec:interaction}

In the case of Interaction-like matrices, we model the diagonal weights $\alpha_{ii}$ as independent and identically distributed (i.i.d.) uniform random variables on the interval $[-d,d]$, \ie:
\begin{equation}\label{diag}
\alpha_{ii}\sim p_D(\theta)=\left\{ \begin{aligned} 
  \frac{1}{2d} \quad \mbox{if} \; \theta\in [-d,d]\\
  0 \quad\mbox{if} \; \; \theta \notin [-d,d]
\end{aligned} \right. .
\end{equation}
The half width of the distribution $d$ is a parameter of our choice, and will have great importance in our study. As regards the off-diagonal elements, we model them as i.i.d. sign-antisymmetric (also called antagonistic) pairs $(\alpha_{ij},\alpha_{ji})$ following a probability density function $p_{\alpha}(u,l)$. For the sake of simplicity, we make the following assumptions on $p_\alpha(u, l)$: the probability for the two sign combinations $(+, -)$ and $(-, +)$ is the same, while the two interaction magnitudes $(|u|, |v|)$ are i.i.d. and independent of the sign. After these assumptions are made, $p_\alpha$ can be expressed as
\begin{equation}\label{eq:palpha}
       p_{\alpha}(u,l)=\frac{1}{2}\tilde{p}(|u|)\tilde{p}(|l|)(\theta(u)\theta(-l)+\theta(-u)\theta(l)),
\end{equation}
where $\tilde{p}(|s|)$ is a normalised probability distribution for the absolute value of the interaction, with support in $[0,+\infty)$. The last thing left to do in order to define $p_\alpha$ is to fix $\tilde{p}$, and we choose to use a uniform distribution:
    \begin{equation}\label{eq:pts}
    \tilde{p}(s)=\left\{ \begin{aligned} 
  \frac{1}{\delta} \quad \mbox{if} \; s\in [0,\delta]\\
  0 \quad\mbox{if} \; \; s\in [0,\delta]
\end{aligned} \right. .
\end{equation}
It is important to notice that multiplying $\textbf{A}$ by a scalar $\beta$ only changes its eigenvalues from $\lambda$ to $\beta\lambda$, and is therefore irrelevant for our study. Accordingly, as explained in Ref.~\cite{valigi2024local}, the absolute magnitude of the diagonal elements should always be compared with the one of the off-diagonal entries, \ie  d must be compared with $\delta$ before drawing meaningful conclusions. For simplicity, we will only vary the value of $d$ and fix $\delta=\sqrt{3}$: this choice enforces the variance 
of $u$ over $p_\alpha$ to be 1, and is coherent with Ref.~\cite{mambuca2022dynamical}.

\subsection{Jacobian-like matrices} \label{sec:jacobian}
The other matrix ensemble we will study describes matrices of the form:
\begin{equation}
\textbf{J}=\textbf{X}(\textbf{A}-\id{N}) \,.
\end{equation}
Here, $\id{N}$ is the identity matrix, $\textbf{A}$ is an Interaction-like matrix as the ones defined above, while $\textbf{X}$ is a diagonal matrix with elements $[\textbf{X}]_{ij}=\delta_{ij}x_i$, where $\delta_{ij}$ is the Kronecker delta. This particular form is chosen to mimic the structure of the Jacobian matrices described in Appendix~\ref{app:linearstability}, while having correlations between the matrix elements that are easier to deal with. The probability distribution of the $x_i$-s is again chosen to be uniform:
\begin{equation}
x_i\sim p_X(\theta)=\left\{ \begin{aligned} 
  \frac{1}{2d_x} \quad \mbox{if} \; \theta\in [1-d_x,1+d_x]\\
  0 \quad\mbox{if} \; \; \theta\notin [1-d_x,1+d_x]
\end{aligned} \right.\,,\quad \mbox{with}\;0\leq d_x\leq 1\,,
\end{equation}
here, $d_x$ is an arbitrary parameter, and its domain is restricted to make the $x_i$-s non-negative: this requirement comes from the ecological setting, in which they represent the population abundances of the species in an ecosystem. 
Similarly to the previous random matrix ensemble, it is the interplay between $d_x$, $d$ and $\delta$ that actually determines the spectral properties of \textbf{J}: again, for simplicity we keep $\delta=\sqrt{3}$ and we take the diagonal elements of \textbf{A} to be equal to $0$, without any randomness, similarly to what has been done in Ref.~\cite{valigi2024local}.


\section{Method for finding the boundary of the spectral support}\label{sec:Algo}

This section is primarily technical and outlines the theoretical framework and algorithmic procedure used to obtain the results presented in Section~\ref{sec:Results}. Readers who are mainly interested in the results may proceed directly to that section.

In Section~\ref{sec:cavity_equations} we briefly recall the cavity approach used to identify the boundary of the spectral bulk in the case of non-Hermitian locally tree-like matrices. This is achieved by analysing the linear stability (see Eqs.~\eqref{trivialcav}, \eqref{epsilon} and \eqref{delta}) of a particular solution of the cavity equations for the hermitised resolvent, commonly referred to as the trivial solution. Our discussion here is intentionally concise and limited to the key ideas. A detailed derivation of the cavity formalism for the spectrum of non-Hermitian random matrices is provided in the Supplemental Material.

In Section~\ref{sec:PopDyn_Algo} we describe the numerical algorithm employed to solve the corresponding self-consistent equations~\eqref{trivialcav}, \eqref{epsilon} and \eqref{delta} determining the linear stability condition.

\subsection{The cavity equations and their trivial solution} \label{sec:cavity_equations}

The main results of this paper have been obtained by studying the behaviour of the leading eigenvalue of the random matrices defined above in the thermodynamic limit. To do so, we have used a technique based on the cavity method and the population dynamics algorithm \cite{mezard2001bethe, rogers2009cavity, metz2019spectral, mambuca2022dynamical}, which we will now briefly describe. At first, we define the limiting spectral distribution of a $N \times N$ matrix $\textbf{M}$ as 
\begin{equation}\label{eq:rho}
    \rho_{\textbf{M}}(z)=\lim_{N\to\infty}\frac{1}{N}\sum_{j=i}^N\delta(z-\lambda_j(\textbf{M})) \, ,\quad z\in\mathbb{C}\, ,
\end{equation}
where $\{\lambda_j(\textbf{M})\}_j{=1,\dots N}$ are the eigenvalues of \textbf{M}. The support of $\rho_{\textbf{M}}$ is defined as
\begin{equation}
    \mathcal{S}_{\textbf{M}}=\{z\in\mathbb{C}:\rho(z) \neq 0\}\,,
\end{equation}
and the leading eigenvalue of $\textbf{M}$, $\lambda^*_j(\textbf{M})$, is defined as the point with the largest real part in the closure of $\mathcal{S}$ (we take the eigenvalue with the largest imaginary part among the candidates for $\lambda_j^*$ whenever this definition leads to ambiguity).

In the case of non-Hermitian matrices, the two-dimensional complex spectrum can not be directly investigated via standard tools from Hermitian random matrix theory. To overcome this, we enlarge the dimensionality by embedding the original $N \times N$ matrix into a $2N \times 2N$ block matrix via a procedure called Hermitisation~\cite{feinberg1997nongaussian, feinberg1997nonhermitian}. Further details are provided in the Supplemental Material. 
In particular, as shown in Refs.~\cite{metz2019spectral, rogers2009cavity}, the evaluation of Eq.~\eqref{eq:rho} in the case of non-Hermitian matrices can be recast as
\begin{equation}
        \rho_\textbf{M}(\lambda)=- \lim_{\eta\to 0^+} \lim_{N\to\infty} \frac{1}{\pi N} \bigg\{\frac{\partial}{\wirtinger}\sum_{i=1}^N\big[\G_{jj}\big]_{11}\bigg|_{z=\lambda}\bigg\} \,,
\end{equation}
where $\eta$ is a regulariser needed to make the computations well defined, $\overline{z}$ is the complex conjugate of $z$, $\frac{\partial}{\wirtinger}$ is the antiholomorphic derivative (also known as Wirtinger derivative), and the $\G_{jj}$-s are $2\times2$ matrices which we refer to as \emph{hermitised resolvent}. In this equation and throughout this section, our convention differs slightly from that of Ref.~\cite{metz2019spectral, mambuca2022dynamical}. 

In the context of locally tree-like matrices~\cite{mezard2001bethe, dembo2010ising, valigi2024local}, \ie, matrices whose associated graphs contain only a small number of short cycles, the hermitised resolvent $\G_{jj}$ can be analysed using the \emph{cavity method}. The central idea of this method is to consider the "cavity graph" obtained by removing a node $l$ and all its incident edges. The resolvent $\G_{jj}$ can then be expressed in terms of cavity quantities $\G^{(l)}_{jj}$ defined on the new graph. On a locally tree-like network, since the number of short cycles is negligible, once we remove $l$ its neighbours either become disconnected or connected through very long paths: for this reason, we can consider the former neighbours of $l$ as independent in the cavity graph. This decorrelation is what allows one to derive closed self-consistency relations for the cavity hermitised resolvents $\G^{(l)}_{jj}$. This construction yields the following recursive relations, commonly referred to as the \emph{cavity equations for the hermitised resolvent}:
\begin{gather}
    \G_{jj}=-\bigg(\z-\M_{jj}+ \sum_{k\in\partial_j}\M_{jk}\G^{(j)}_{kk}\M_{kj}\bigg)^{-1},\label{diagGtrees}\\
    \G^{(l)}_{jj}=-\bigg(\z-\M_{jj}+ \sum_{k\in\partial_j	\setminus\{l\}}\M_{jk}\G^{(j)}_{kk}\M_{kj}\bigg)^{-1}\,,\label{finalgcav}
\end{gather}
where $\partial_j$ denotes the set of neighbours of node $j$ in the interaction network, the superscript $^{(j)}$ refers to quantities computed on the cavity graph obtained by removing node $j$, and we have also introduced the following $2 \times 2$ matrices:
\begin{equation}\label{Mzdef}
    \M_{jk}= \begin{pmatrix}
    [\textbf{M}]_{jk} & 0 \\
    0 & \overline{[\textbf{M}]_{kj}} \end{pmatrix}
     \quad , \quad  \z= \begin{pmatrix}
    z & i\eta \\
    i\eta & \overline{z} \end{pmatrix}.
\end{equation}


When $\eta=0$, the cavity equations admit a diagonal solution, called the \emph{trivial solution}, which is only valid outside the spectrum~\cite{metz2019spectral, mambuca2022dynamical}:
\begin{equation}
    \G^0_{jj}=
    \begin{pmatrix}
        g_j & 0\\
        0 & \overline{g}_j
    \end{pmatrix}, \quad \G^{0,(l)}_{jj}=
    \begin{pmatrix}
        g^{(l)}_j & 0\\
        0 & \overline{g}^{(l)}_j
    \end{pmatrix}, \label{trivialG}
\end{equation}
where the superscript $^0$ is used to distinguish the trivial solution from the other possible ones, and
\begin{gather}
    g_{j}=\frac{-1}{z-[\textbf{M}]_{jj}+ \displaystyle{\sum_{k\in\partial_j}}[\textbf{M}]_{jk}g^{(j)}_{k}[\textbf{M}]_{kj}}\label{trivial},\\
    g^{(l)}_{j}=\frac{-1}{z-[\textbf{M}]_{jj}+ \displaystyle\sum_{k\in\partial_j	\setminus\{l\}}[\textbf{M}]_{jk}g^{(j)}_{k}[\textbf{M}]_{kj}}\label{trivialcav}.
\end{gather}
Another non diagonal solution must exist in the same limit inside the spectrum.
To study whether a point is inside of the spectrum or not, still avoiding to evaluate the full solution, one could perturb \eqref{trivialG} by adding small, off diagonal elements to each of the $\G^{0,(l)}_{jj}$, thus obtaining
\begin{equation}
    \G^{p,(l)}_{jj}=
    \begin{pmatrix}
        g^{(l)}_j & \delta_j^{(l)}\\
        \epsilon_j^{(l)} & \overline{g_j}^{(l)}
    \end{pmatrix},
\end{equation}
where $\delta_j^{(l)}$ and $\epsilon_j^{(l)}$ are both small real numbers, and the superscript $^p$ is used to identify the perturbed trivial solution. After that, a linear stability analysis can be employed to check if the trivial solution is stable or not: inside of the spectrum, where it is not defined, we expect the perturbations to diverge when iteratively applying the cavity equations, whereas we expect them to vanish outside of $\mathcal{S}$. In order to check the linear stability of the solution we expand Eq.~\eqref{finalgcav} at first order in the perturbations $\epsilon$ and $\delta$, getting
\begin{gather}
\delta_{j}^{(l)}=\bigg|g_j^{(l)}\bigg|^{2}\displaystyle\sum_{k\in \partial_j\setminus\{l\}}\delta_{k}^{(j)} |[\textbf{M}]_{jk}|^2, \label{epsilon}\\
\epsilon_{j}^{(l)}=\bigg|g_j^{(l)}\bigg|^{2}\displaystyle\sum_{k\in \partial_j\setminus\{l\}}\epsilon_{k}^{(j)} |[\textbf{M}]_{kj}|^2,\label{delta}
\end{gather}
where the $g_j^{(l)}$-s satisfy Eq.~\eqref{trivialcav}. The behaviour of $\epsilon_{j}^{(l)}$ and $\delta_{j}^{(l)}$ can then be studied with these equations to analyse the stability of the trivial solution, and that's where the population dynamics algorithm, discussed in Section~\ref{sec:PopDyn_Algo} and Appendix~\ref{app:PopDyn}, comes into play~\cite{metz2019spectral}.

Before proceeding with the description of the algorithm, it is important to stress that, while the trivial solution typically loses stability in the bulk, it remains stable in certain components of the spectrum, as the singular ones~\cite{metz2019spectral}: for this reason, the linear stability analysis of the trivial solution does not generally capture the presence of these regions. Although, to the best of our knowledge, no rigorous results are available on this matter, previous studies suggest that this procedure correctly identifies the bulk for the ensembles considered here~\cite{mambuca2022dynamical}: in this case, the Population Dynamics algorithm gives us the possibility to estimate the position of the bulk's boundary, thus providing a great advantage in the identification of the different spectral phases. However, as we are going to show in the following sections, we observe that this approach is affected by a systematic underestimation of the real-axis segment that emerges from the bulk in the ensembles under study. 
We attribute this underestimation to the fact that the eigenvalues in this segment belong to the singular component of the spectrum; consequently, a different approach is needed to accurately estimate the extent of this emerging segment. Previous results~\cite{valigi2024local} indicate that a reliable proxy for it is given by the support of the distribution of diagonal elements. For this reason, as discussed in more detail in Appendix~\ref{app:emergence} and~\ref{app:Disc}, whenever a real-axis segment emerges from the bulk we will use the parameters $d$ and $d_x$, introduced in Section~\ref{sec:Model}, as proxies for its extent in Interaction-like and Jacobian-like matrices, respectively. As additional confirmation of the reliability of our method, in the next section we are going to systematically compare our population dynamics predictions with direct diagonalisation results.

\subsection{Distributional equations for the quantities of interest} \label{sec:PopDyn_Algo}

A way to deal with Eqs.~\eqref{trivialcav}, \eqref{epsilon} and \eqref{delta} is to recast them as distributional equations in the thermodynamic limit, and study the ensemble averages of the quantities of interest \cite{metz2019spectral, neri2016eigenvalue,mambuca2022dynamical}. In order to do so, we consider that the empirical joint distribution of the diagonal and off-diagonal elements of $\G_{jj}^{p,(l)}$, which will be called fields from now on:
\begin{equation}\label{empirical}
    Q_\textbf{M}(g,h,h')=\frac{1}{cN}\sum_{j=1}^N\sum_{l\in\partial_j}\delta(g-g_{j}^{(l)})\delta(h-\epsilon_{j}^{(l)})\delta(h'-\delta_{j}^{(l)}) \, .
\end{equation}
Note that, as follows from Eq.~\eqref{trivialcav}, $Q_\textbf{M}$ depends on $z$ but keep it implicit to make the notation lighter. If $Q_\textbf{M}$ is self averaging, as $N\to\infty$ it will tend to its ensemble average $Q(g,h,h')$, and lose its dependence on the specific realization of $\textbf{M}$. Furthermore, all neighbourhoods $\{\partial_j\}$ of the interaction network will become statistically equivalent realizations of the same random variables: this means that we can forget the notion of the graph and just focus on single nodes and their $K'$ neighbours, where $K'$ is a random variable distributed according to $p_{res}$, the residual degree distribution of the graph:
\begin{equation}\label{residualappendix}
    p_{res}(k)=\frac{(k+1)p_{deg}(k+1)}{\langle k \rangle_{p_{deg}}}\,.
\end{equation}
In this framework, we can always drop the index referring to the central node, and just label its neighbours with a single index. Bearing this in mind, we can write a general equation for $Q$:
\begin{gather}
    Q(g,h,h')=\sum_{k=0}^\infty p_{res}(k)\int\prod_{i=1}^kd g_id h_id h'_id \tilde{u}_id \tilde{l}_id\tilde{\theta}\,\hat{Q}(\{g_i,h_i,h'_i,\tilde{u}_i,\tilde{l}_i\}_{i=1,...,N},\tilde{\theta})\times\nonumber\\\times\delta\bigg(h-|g|^{2}\displaystyle\sum_{i=1}^{k}h_i |{\tilde{u}_i}|^2\bigg)\delta\bigg( g+\frac{1}{z-\tilde{\theta}+ \sum_{i=1}^{k}{\tilde{l}_i}g_{i}{\tilde{u}_i}}\bigg)\delta\bigg(h'-|g|^{2}\displaystyle\sum_{i=1}^{k}h'_i |{\tilde{l}_i}|^2\bigg). \label{PopDynGen}
\end{gather}
where $d g$,$d h$ and $d h'$ denote bidimensional Lebesgue measures in $\mathbb{C}$, while similarly\footnote{Here, we add a tilde to highlight that the matrix elements do not necessarily belong to Interaction-like matrices} to Section~\ref{sec:Model}, $\tilde{u_i},\tilde{l_i},\tilde{\theta}$ represent the random variables corresponding to the matrix elements $[\textbf{M}]_{ij}$$, [\textbf{M}]_{ji}$, $[\textbf{M}]_{jj}$. In addition to that, we have used $\hat{Q}$ to denote the joint distribution of the fields defined on the nodes of the neighbourhood $(g_i,h_i,h'_i)$, and the matrix elements $(\tilde{u}_i,\tilde{l}_i,\tilde{\theta})$. Equation~\eqref{PopDynGen} is, in a sense, just a change of variables formula: the fields (g,h,h') on a node depend on the ones on its neighbours and the matrix elements through Eqs.~\eqref{trivialcav}, \eqref{epsilon} and \eqref{delta}, and \eqref{PopDynGen} is making this explicit in a distributional sense. Expressing the relation between $\hat{Q}$, $Q$, and the distribution of the matrix elements will allow us to solve \eqref{PopDynGen} in an iterative, self consistent way, through the population dynamics algorithm \cite{mezard2001bethe}.

\subsubsection{Distributional equations in the Interaction-like case}
In the case of Interaction-like matrices, defined in Section \ref{sec:interaction}, we can write:
\begin{equation}
    \hat{Q}(\{g_i,h_i,h'_i,\tilde{u}_i,\tilde{l}_i\}_{i=1,...,N},\tilde{\theta})=p_D(\theta)\prod_iQ(g_i,h_i,h_i')p_{\alpha}(u_i,l_i)
\end{equation}
this happens because, as already stated, the fields in the original graph are computed in absence of the central node $j$: the presence of this cavity completely decouples all the quantities in play, and makes the probability distributions factorize. Equation \eqref{PopDynGen} now becomes:
\begin{gather}
    Q(g,h,h')=\sum_{k=0}^\infty p_{res}(k)\int\prod_{i=1}^k d  g_i d  h_i d  h'_iQ(g_i,h_i,h'_i)\int\prod_{i=1}^k d  u_i d  l_i p_\alpha(u_i,l_i)\int d \theta p_D(\theta)\times\nonumber\\\times\delta\bigg(h-|g|^{2}\displaystyle\sum_{i=1}^{k}h_i |{u_i}|^2\bigg)\delta\bigg( g+\frac{1}{z-\theta+ \sum_{i=1}^{k}{l_i}g_{i}{u_i}}\bigg)\delta\bigg(h'-|g|^{2}\displaystyle\sum_{i=1}^{k}h'_i |{l_i}|^2\bigg), \label{PopDynInt}
\end{gather}

\subsubsection{Distributional equation in the Jacobian-like case}

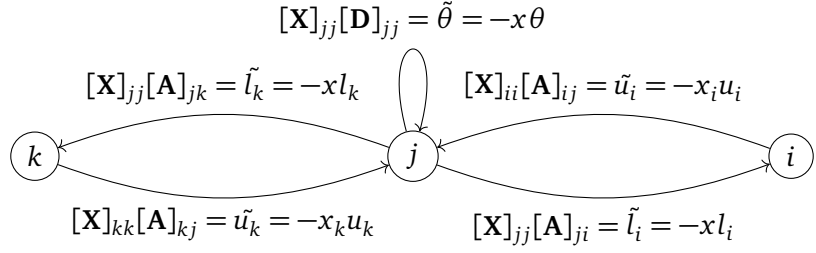
\begin{figure}
    \centering
    \begin{tikzpicture}[node distance={50mm}, main/.style = {draw, circle, inner sep=3pt}] 
\node[main]  (j) at (0,1) {$j$}; 
\node[main] [right of=j] (i) {$i$};
\node[main] [left of=j] (k) {$k$};

\draw[->] (j) edge[bend right=20] node[below,sloped] {$[\textbf{X}]_{jj}[\textbf{A}]_{ji}=\tilde{l_i}=-xl_i$} (i) ;
\draw[->] (j) edge [loop above,min distance=15mm,looseness=10] node {$[\textbf{X}]_{jj}[\textbf{D}]_{jj}=\tilde{\theta}=-x\theta$} (j);
\draw[->] (i) edge[bend right=20] node[above,sloped] {$[\textbf{X}]_{ii}[\textbf{A}]_{ij}=\tilde{u_i}=-x_iu_i$} (j) ;
\draw[->] (j) edge[bend right=20] node[above,sloped] {$[\textbf{X}]_{jj}[\textbf{A}]_{jk}=\tilde{l_k}=-xl_k$} (k) ;
\draw[->] (k) edge[bend right=20] node[below,sloped] {$[\textbf{X}]_{kk}[\textbf{A}]_{kj}=\tilde{u_k}=-x_ku_k$} (j) ;
\end{tikzpicture}
    \caption{Explicative diagram showing the differences between the matrix elements involved in the cavity equations. We already show how we will apply Eqs. \eqref{utildeJ}-\eqref{thetatildeJ}. }
    \label{fig:JacobianElements}
\end{figure}
In this case we recall that we are studying matrices of the form $\textbf{J}=-\textbf{XA}$. The presence of $\textbf{X}$ introduces some correlations that were absent before, because the population abundances $\{x_i\}$ multiply all the elements of the same row. As shown in \figref{fig:JacobianElements}, the effect of neighbouring sites $i$ on $j$ carry the contribution of their population abundances $x_i$, while we must count the population abundance $x$ on the effect of $j$ on the neighbouring $i$.  This fact induces the following definitions:
\begin{gather} 
    \tilde{u_i}=-x_iu_i,\label{utildeJ}\\
    \tilde{l_i}=-xl_i,\\
    \tilde{\theta}=-x\theta,\label{thetatildeJ}
\end{gather}
where $u_i$, $\theta$ and  $l_i$ represent, as always, realisations of the entries of the interaction matrix, while $x_i$ and $x$ are i.i.d. random variables drawn from $p_X$, and represent the diagonal elements of \textbf{X}. Taking into account the correlations between the population abundances and the fields, we now have:
 \begin{gather} 
    \hat{Q}(\{g_i,h_i,h'_i,\tilde{u}_i,\tilde{l}_i\}_{i=1,...,N},\tilde{\theta})=p_D(\theta)p_X(x)\prod_iQ(g_i,h_i,h_i'|x_i)p_X({x_i)}p_{\alpha}(u_i,l_i)
\\
    Q(g,h,h')=\sum_{k=0}^\infty p_{res}(k)
    \int\prod_{i=1}^k d  x_i p_X(x_i)
    \int\prod_{i=1}^k d  g_i d  h_i d  h'_iQ(g_i,h_i,h'_i|x_i)\times\nonumber\\
    \times\int\prod_{i=1}^k d  u_i d  l_i p_\alpha(u_i,l_i)
    \int d \theta  d  x p_D(\theta)p_X(x)\times\nonumber\\
    \times\delta\bigg(h-|g|^{2}\displaystyle\sum_{i=1}^{k}h_i x_i^2|{u_i}|^2\bigg)
    \delta\bigg( g+\frac{1}{z+x\theta+ x\sum_{i=1}^{k}{l_i}g_{i}x_i{u_i}}\bigg)
    \delta\bigg(h'-|g|^{2}x^2\displaystyle\sum_{i=1}^{k}h'_i |{l_i}|^2\bigg)\,,\label{PopDynJac}
\end{gather}
where we have made an abuse of notation, using $Q$ to indicate also the conditional distribution of the fields with respect to the abundances. We stress the fact that neglecting the correlations with the fields, \emph{i.e.,} using $Q(g,h,h')$ instead of $Q(g,h,h'|x_i)$ in the integral, would lead to significant errors in the estimation of the spectral boundary. In the following, in order to reduce the number of relevant parameters, we impose that the diagonal elements of $\textbf{A}$ are all equal to $1$, as anticipated in Section~\ref{sec:jacobian}:
\begin{equation}
p_D(\theta)=\delta(\theta-1)\,,
\end{equation}

\subsubsection{Numerical validation of the equations}\label{sec:validation}
\begin{figure}[ht]
        \centering
        
        \begin{subfigure}[b]{\textwidth}
         \centering
         \includegraphics[width=.9\textwidth]{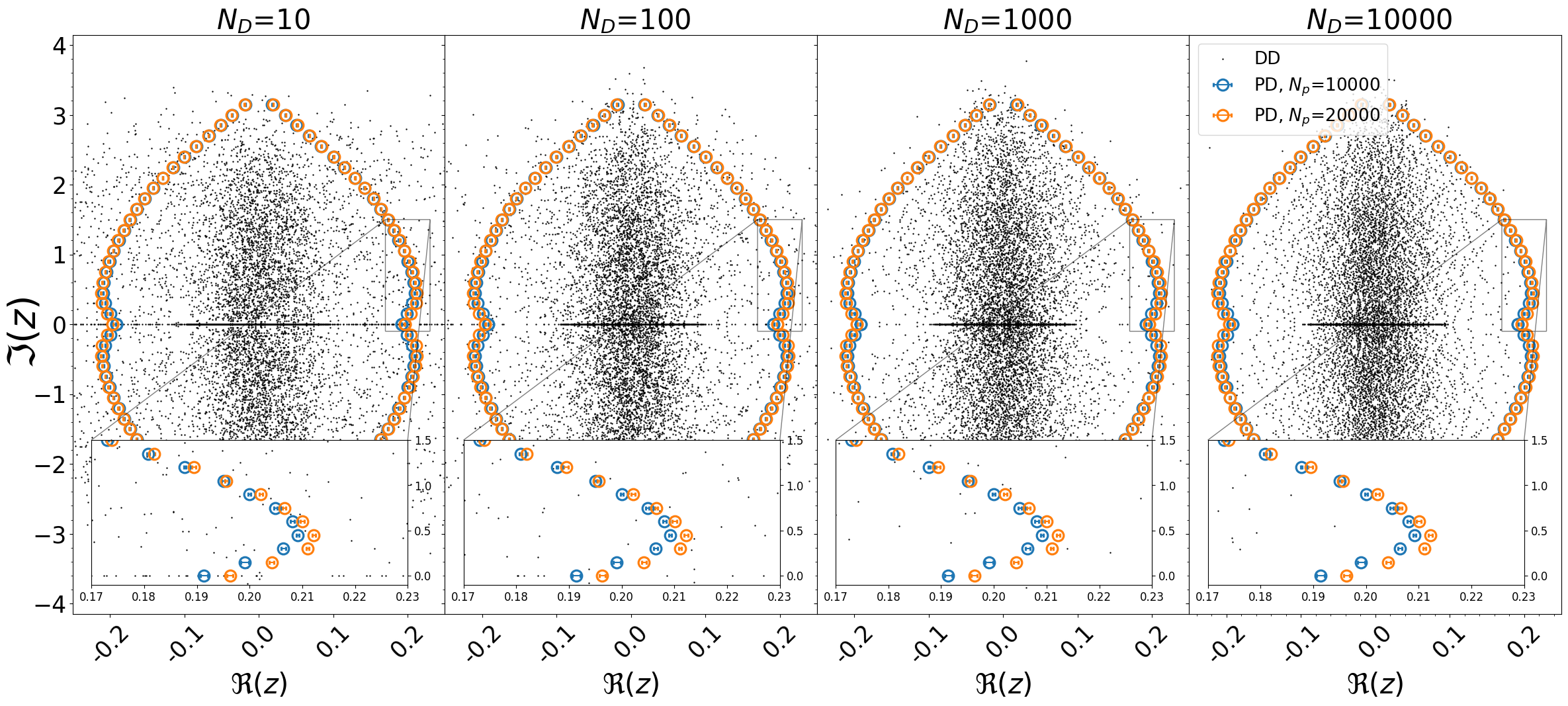}
         \caption{Interaction-like matrix $c=2.1$, $d=0.1$}
        \label{fig:InteractionComparisons}
     \end{subfigure}
     \begin{subfigure}[b]{\textwidth}
         \centering
         \includegraphics[width=.9\textwidth]{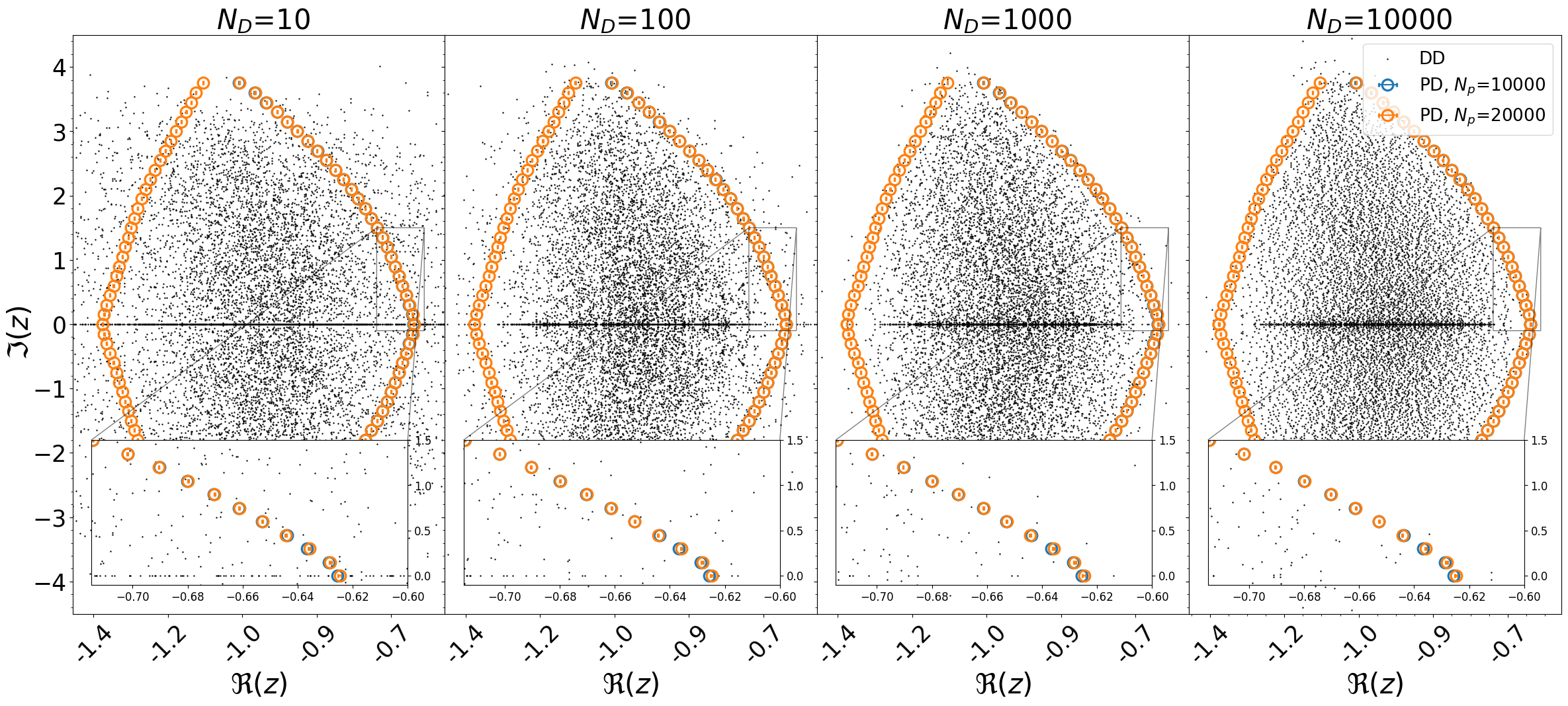}
         \caption{Jacobian-like matrix, $c=3$, $d_x=0.3$}
        \label{fig:JacobianComparisons}
     \end{subfigure}
    \caption{Results of the Population Dynamics algorithm (PD) and Direct Diagonalisation (DD) method, for one instance of Interaction-like matrix (above) and one of Jacobian-like matrix (below). We include a zoom of the highlighted region for better visualisation. The matrices belong to the ensembles discussed in Sec. \ref{sec:interaction} and Sec. \ref{sec:jacobian}, and the relevant parameters can be found in the picture. At lower sizes of the matrices ($N_D$), we display the eigenvalues of different matrices belonging to the same ensemble, so that the number of black dots in each plot is always equal to $10000$.}
    \label{fig:FirstComparison}
\end{figure}

\begin{figure}[ht]
    \centering
    \includegraphics[width=.9\textwidth]{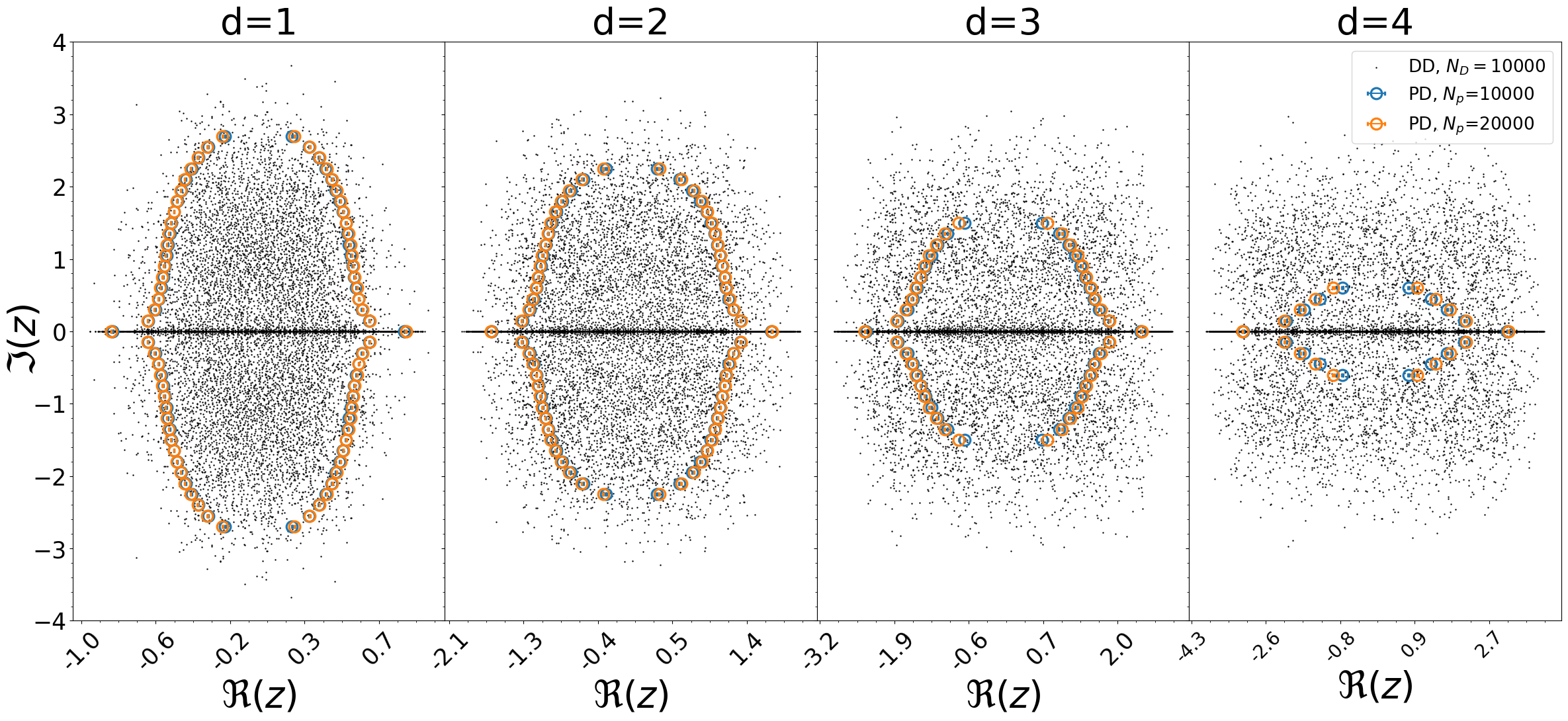}
    \caption{An example of how our algorithm can fail for strong disorder in the interaction-like case ($c=2.1$). As in the previous Figure, the matrices belong to the ensemble discussed in Sec. \ref{sec:interaction}, and the relevant parameters are displayed in the picture.}
    \label{fig:PopDynFail}
\end{figure}

As already stated, and as discussed in detail in Appendix \ref{app:PopDyn}, we can study Eqs. \eqref{PopDynInt} and \eqref{PopDynJac} iteratively, through the Population Dynamics algorithm. In particular, the algorithm allows us to compute the ensemble averages of $\langle|h|\rangle$ and $\langle|h'|\rangle$, which in turn give us a criterion to check whether a point $z$ is in the spectrum or not. Through bisection, we can then find the border of the spectral support $\mathcal{S}$, and compare the results with the eigenvalues obtained by direct diagonalisation of large matrices belonging to the considered ensembles, as done in \figref{fig:FirstComparison}. As we can see, the accordance becomes better as the size of the matrices increases: this could be expected by the fact that our procedure is built to obtain the boundary of the spectrum in the limit $N\to\infty$. Moreover, we notice that, especially around the real axis, the borders we find slightly depend on the number of samples $N_p$ (see insets of top panels in \figref{fig:FirstComparison}) that we use to approximate the studied distributions, which highlights the presence of some finite size effects also in the Population Dynamics algorithm. On the other hand, in \figref{fig:PopDynFail}, we show examples of how our algorithm is not able to correctly estimate the boundary when the diagonal disorder is large. Specifically, we have observed that the strength of diagonal disorder $d$ at which Population Dynamics begins to deviate from direct diagonalisation shifts to lower values as the connectivity $c$ is reduced. The clear relation between these two parameters, however, remains to be determined. Similar issues are also observed in the Jacobian-like case. 
From what shown in \figref{fig:PopDynFail}, it is evident that the observed discrepancies can not be attributed solely to finite-size effects, but rather to the fact that the trivial solution does not lose stability in correspondence of certain localised components of the spectrum, as anticipated at the end of Section~\ref{sec:cavity_equations}. Further details on the origin of this discrepancy, together with a possible resolution, will be discussed in a future paper currently in preparation. Note that the main results of this work are obtained in a regime where these discrepancies are essentially negligible. The only visible errors at relatively low disorders are, as already anticipated, the ones on the real axis spike, as shown in the fourth panel of both \figref{fig:InteractionExamples} and \figref{fig:JacobianExamples}. However, as done in Appendix~\ref{app:Disc}, we can use the the support of the diagonal elements' distribution to locate it, thus mitigating the effect that these errors have on the phase diagram.

\section{Main results/Phase diagrams}\label{sec:Results}

\begin{figure}[ht]
    \centering
    \begin{subfigure}[b]{\textwidth}
         \centering \includegraphics[width=.9\textwidth]{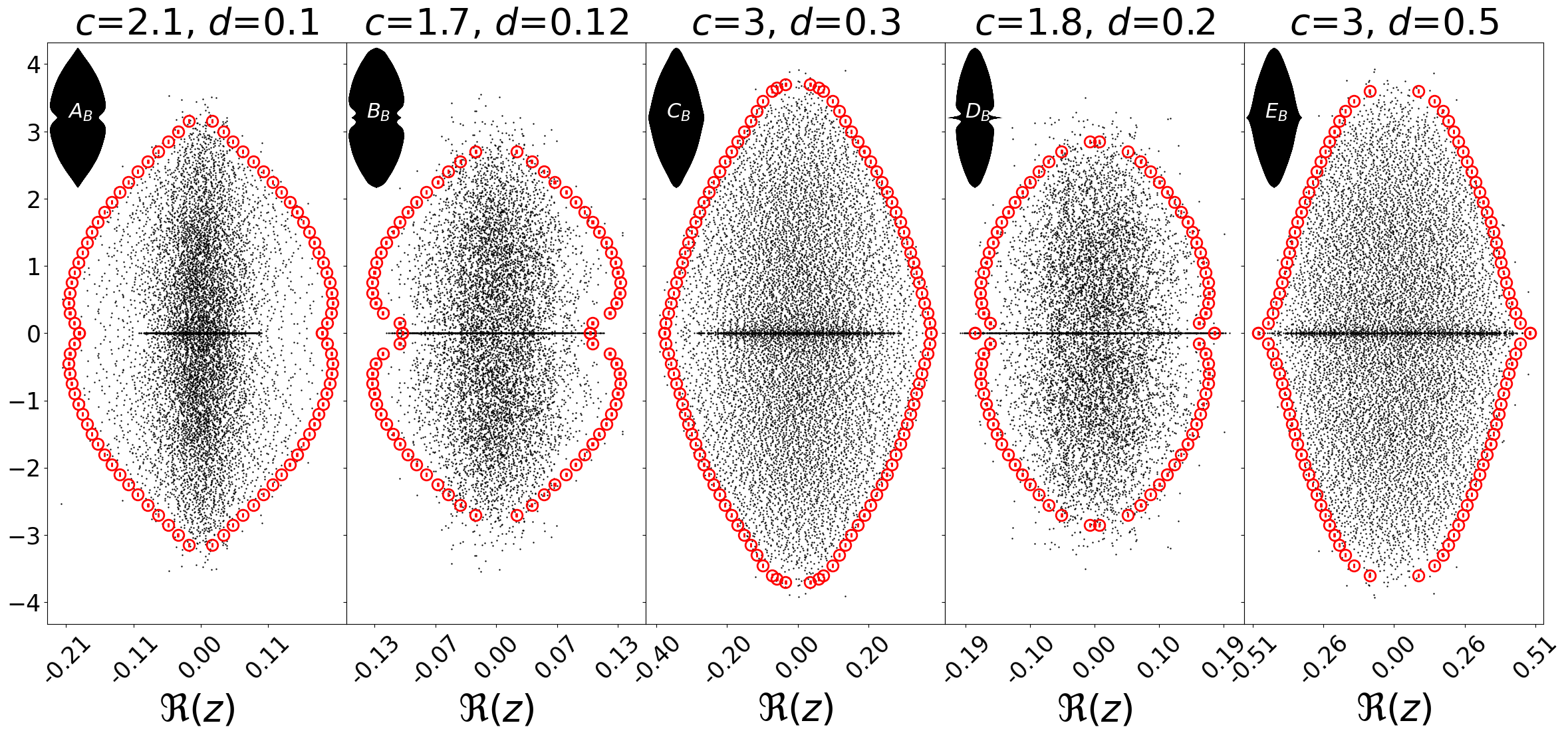}
         \caption{Interaction-like matrices}\label{fig:InteractionExamples}
     \end{subfigure}
     
     \vspace{0.5cm}
     
     \begin{subfigure}[b]{\textwidth}
         \centering \includegraphics[width=.9\textwidth]{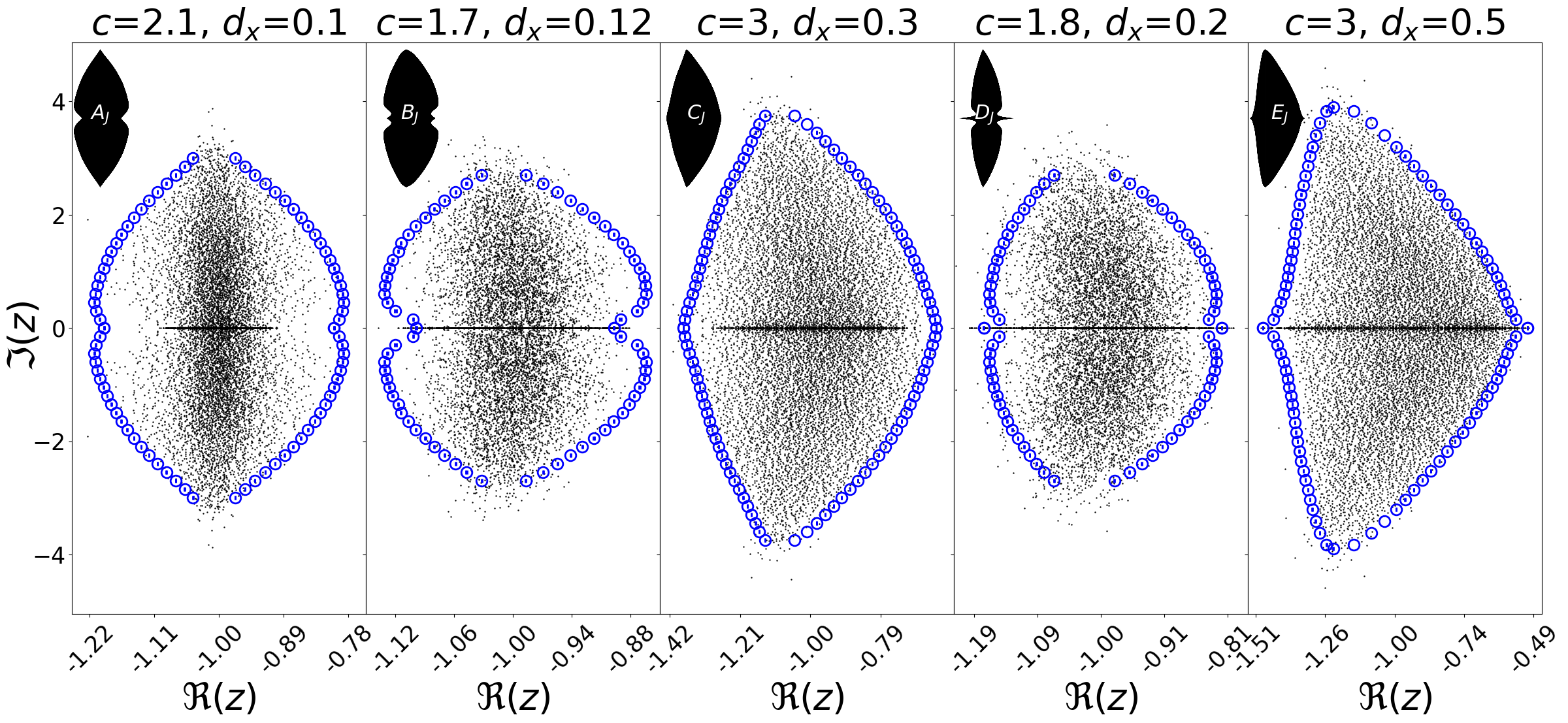}\caption{Jacobian-like matrices} \label{fig:JacobianExamples}
     \end{subfigure}
    \caption{Examples of all of the behaviours observed when changing the parameters of our model. The points are computed using $N_p=20000$ and averaging over $N_m=10$ realisations. The direct diagonalisation eigenvalues (black dots) are computed by diagonalizing matrices of linear size $N_D$ belonging to the ensembles described in Sec. \ref{sec:interaction} and Sec. \ref{sec:jacobian}. All the other relevant parameters are displayed in the picture. Notice how the underestimation of the real axis segment makes it hard for Population Dynamics to discern that the emergence transition has already occurred for the numerical examples at c=1.7.}
    \label{fig:Examples}
\end{figure}

\begin{figure}[!ht]
     \centering
     \begin{subfigure}[b]{.9\textwidth}
         \centering
         \includegraphics[width=\textwidth]{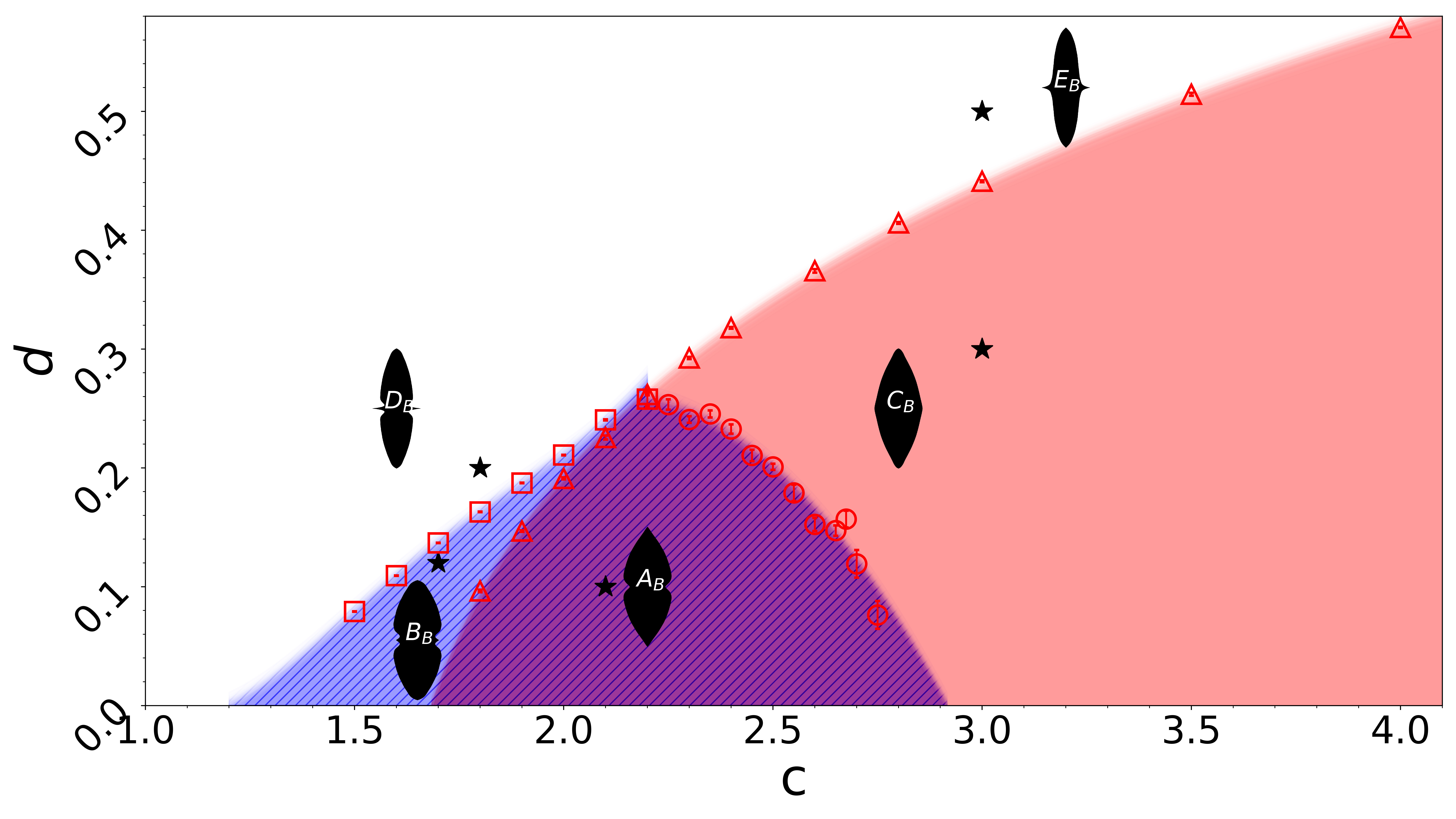}
         \caption{\textbf{Interaction-like matrices.}}
     \end{subfigure}
     
     \begin{subfigure}[b]{\textwidth}
         \centering
         \includegraphics[width=.9\textwidth]{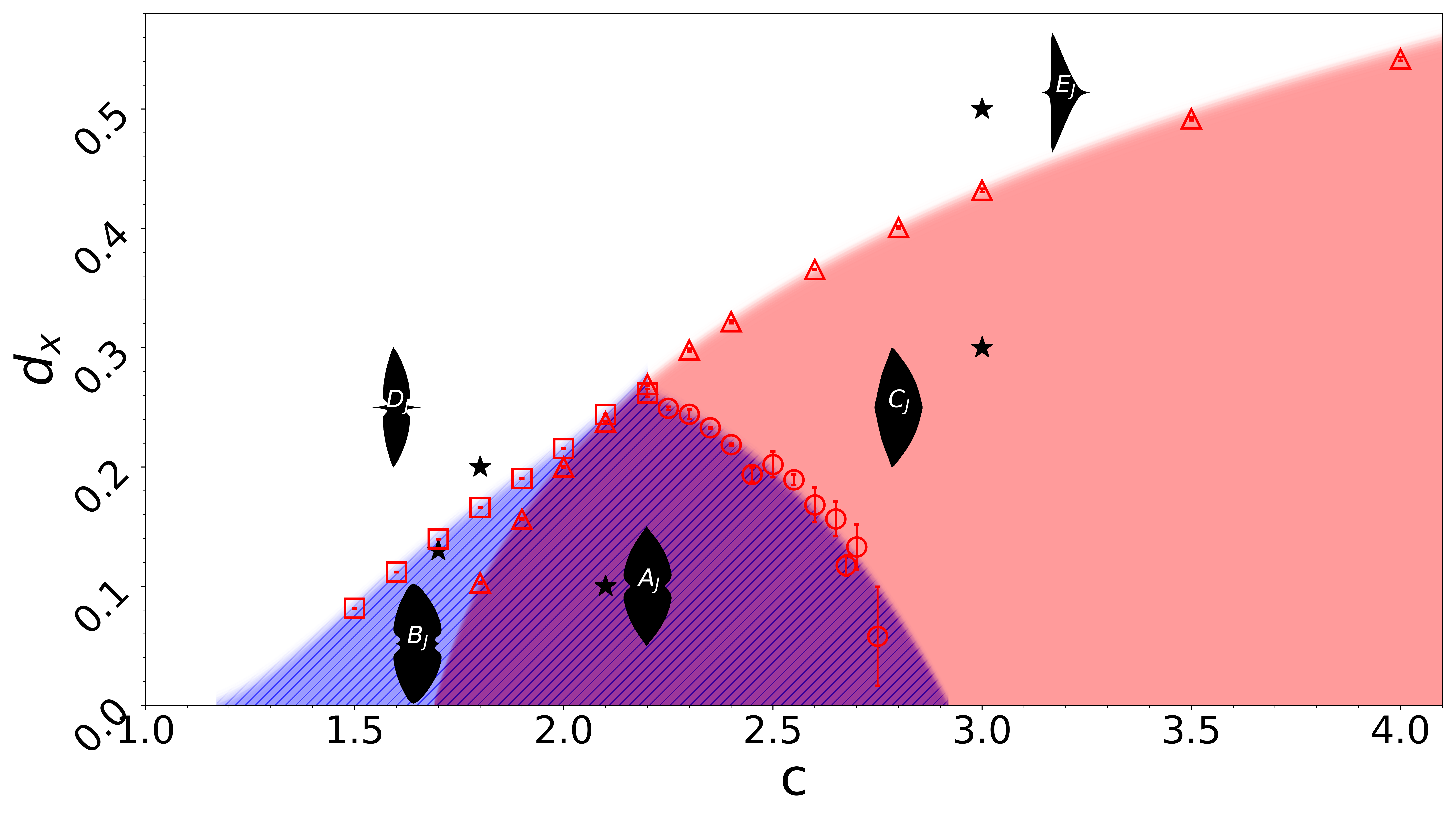}
         \caption{\textbf{Jacobian-like matrices.}}
        
     \end{subfigure}
        \caption{Phase diagrams of both matrix ensembles. The shaded areas highlight different regions in the space of the parameters, where the spectra behave differently, and each of them is labelled with a different letter. The hatched regions are the ones where the leading eigenvalue is complex. Star-shaped markers are placed in correspondence to the examples presented in Fig.\ref{fig:Examples}. Triangular markers represent the emergence transition points, square markers the discontinuous transition points and circular markers the continuous transition ones, all of which have been found as described in Appendices \ref{app:emergence}, \ref{app:Disc} and \ref{app:Cont} respectively. We stress that the lines separating the different phases must only be intended as a guide for the eye. Notice that the silhouettes are not in scale and are to be interpreted as shortcuts for the spectral phases presented in Fig.~\ref{fig:Examples}.}
        \label{fig:PhaseAreas2}
\end{figure}

Using the algorithms described in Appendix \ref{app:PopDyn} we have observed that all the spectra under consideration have finite width, similarly to what found in Ref.~\cite{mambuca2022dynamical}. As shown in \figref{fig:Examples}, the shape of the spectral support, which influences the properties of the leading eigenvalue, depends on both the connectivity $c$ of the interaction network, and the level of diagonal disorder ($d$ or $d_x$). This phenomenon can be summarized in the following regimes:
\begin{itemize}
    \item \textbf{Low values of $c$}: a significant indentation, which we also refer to as reentrance, is present near the real axis for low diagonal disorder. If the latter is increased, a swelling of $\mathcal{S}$ is observed (see first panel of \figref{fig:InteractionExamples} and \figref{fig:JacobianExamples})\\\, and eventually a spike is formed on the real axis (see second panel of \figref{fig:InteractionExamples} and \figref{fig:JacobianExamples}), a phenomenon that we call \textbf{emergence transition}. When $d$ ($d_x$) is large enough, the spike surpasses the rest of the spectrum while the indentation is still present (see fourth panel of \figref{fig:InteractionExamples} and \figref{fig:JacobianExamples}): this produces a jump of the leading eigenvalue's imaginary part, which we call \textbf{discontinuous transition} in $d$ or $d_x$. As $d$ ($d_x$) becomes larger and larger the whole spectrum deforms to keep up with the increasingly long spike, and at some point the reentrance is completely lost.
    \item \textbf{Intermediate values of $c$}: this time, the indentation at $d=0$ is smaller, and increasing the diagonal disorder produces a swelling of the spectrum that is sufficient to absorb the reentrance before the emergence of the spike: this translates to the fact that $\Im(\lambda^*)$ goes to zero smoothly, a phenomenon we call \textbf{continuous transition} in $d$ ($d_x$). After that, the spectrum continues to increase in size, and at some point the emergence transition takes place, so that a spike is formed on the real axis.
    \item \textbf{High values of $c$}: In this case there is no indentation to begin with, hence we only observe the swelling of the spectrum and the emergence transition (see third and fifth panel of \figref{fig:InteractionExamples} and \figref{fig:JacobianExamples}).
\end{itemize}
These phenomena are coherent to the reentrance effect at low values of $c$ described in Ref.~\cite{mambuca2022dynamical}, and with the spectra obtained through direct diagonalisation in Ref.~\cite{valigi2024local}. In particular, the formation of the spike is due to the accumulation of eigenvalues on a segment of the real axis corresponding to the distribution of the matrices' diagonal elements, which can be observed in \figref{fig:Examples}: if the segment is small, the disorder just produces a swelling of $\mathcal{S}$, whereas if it is large enough, it protrudes from the rest of the spectrum as a spike. Our observations allow us to identify five different regions in the parameters space for both Interaction like and Jacobian-like matrices. These regions are divided by the emergence transition, continuous transition and discontinuous transition lines, which we have identified with the methods discussed in Appendices \ref{app:emergence} to \ref{app:Cont} and depicted in \figref{fig:PhaseAreas2}. In summary, the regions we have identified are the following:
\begin{itemize}
    \item \textbf{Regions $A_B$ and $A_J$}, highlighted in \textbf{purple} in both pictures: here, no transition has occurred yet, thus the reentrance effect is present, and there are no spikes protruding from the boundary. The leading eigenvalue of the matrices in this regions is imaginary.  We highlight the fact that if we were to increase the diagonal disorder's strength, not all matrices belonging to this region would behave the same: in fact, there is a threshold value of the connectivity\footnote{In principle this value could be different in the two matrix ensembles, but we have found it to be very similar between the two cases.} $c_s\approx2.2$ above which $\Im(\lambda^*)$ will undergo a continuous transition in $d$ or $d_x$, and a discontinuous one otherwise.
    \item \textbf{Regions $B_B$ and $B_J$}, highlighted in \textbf{blue} in both pictures: here the emergence transition has already happened, meaning that the spectra of the matrices belonging to these regions display a spike on the real axis. The spike spans the interval on which the diagonal elements are distributed, but it is still too short to contain the leading eigenvalue. As a result, $\lambda^*$  has a nonzero imaginary part.
    \item \textbf{Regions $C_B$ and $C_J$}, highlighted in \textbf{pale red} in both pictures: here there is no indentation, as the continuous transition in either $d$ ($d_x$) or $c$ has already occurred. On the other hand, the emergence transition has not taken place yet, meaning that the spectra do not display any spike on the real axis.
    \item \textbf{Regions $D_B$ and $D_J$}: here the discontinuous transition has already taken place, but a small reentrance effect still survives: the spectra of the matrices belonging to these regions have a real leading eigenvalue, but still display the change in concavity caused by the indentation.
    \item \textbf{Regions $E_B$ and $E_J$}: here the matrices only display the spike on the real axis, without any indentation whatsoever. This phenomenon can either be caused by the fact that the reentrance effect was not present from the beginning, or by a diagonal disorder so strong that absorbs any pre-existing indentation through the swelling of the border. Both of these cases are displayed in Figure \figref{fig:PhaseAreas2}
\end{itemize}

The separating line between the $D$ and $E$ regions does not correspond to any of the transitions we have studied, thus we do not represent it. 
Its location should be obtained detecting the change of concavity in the boundary, which entails the technical challenges described in Appendix~\ref{app:Cont}. Furthermore, this transition does not affect the leading eigenvalue, which belongs to the segment on the real axis across both phases. 
However, we know that such a line must exist: in fact, given a strong enough diagonal disorder, all spectra behave as in $E_B$ and $E_J$. We stress the fact that the curves delimiting all the regions are to be intended as approximations extrapolated from the data points and are not exact estimates. Finally, we highlight how the phase diagrams of the Jacobian-like matrices clearly resemble the ones of the Interaction-like matrices from which they are obtained. This is probably caused by the specific form we have chosen for our Jacobian-like matrices, which becomes an interaction one as $d_x\to0$, and this means that the shape of the spectra in the two ensembles are quite similar for low disorder.

\section{Discussion} \label{sec:discussion}

In this work, we have investigated the spectral properties of sparse, antagonistic random matrices, delineating a phase diagram characterised by five distinct phases. Motivated by the dynamics of complex systems, particularly complex ecosystems, we focused on two distinct classes of matrices: Interaction-like and Jacobian-like matrices. By building on the cavity method on locally tree-like random graphs~\cite{metz2019spectral}, we derived distributional equations that can be numerically solved by means of the Population Dynamics algorithm to locate the boundary of the spectral support for these sparse, non-Hermitian, random matrix ensembles.

Our theoretical framework, validated by extensive numerical diagonalisation, has the advantage of providing, almost always with high precision, the boundary of the spectral support, thereby shedding light on how the interplay between sparsity and diagonal disorder shapes the spectral boundary in both Interaction-like and Jacobian-like ensembles. Building on the previous observation of spectral reentrances in purely antagonistic systems~\cite{mambuca2022dynamical, valigi2024local}, we investigate how these concave features are modified by the presence of diagonal disorder. Our study reveals that increasing disorder induces a "swelling" of the spectral support and the emergence of an eigenvalue segment on the real axis, which gives rise to a number of peculiar shapes of the spectral support, going well beyond the classical elliptical law and even beyond what discussed in Refs.~\cite{mambuca2022dynamical, valigi2024local}. 

The behaviour we find at low values of the diagonal disorder $d$ is emblematic of how richer the disordered case is with respect to the previously studied ones: at $d=0$ when varying the connectivity, only one continuous transition was observed in Ref.~\cite{mambuca2022dynamical}, while we observe three distinct transitions at fixed low but non-zero disorder: at very low connectivity, the leading eigenvalue is real and determined by the real-axis segment (Phase $D$ in Fig.~\ref{fig:PhaseAreas2}, non-oscillatory). As the connectivity grows, the spectral support widens, inducing a discontinuous transition where the leading eigenvalue becomes complex (Phases $B$ and $A$ in Fig.~\ref{fig:PhaseAreas2}, oscillatory). Further increasing the connectivity progressively reshapes the spectral boundary, until the reentrances vanish, leading to a continuous transition where the leading eigenvalue becomes real again (Phase $C$ in Fig.~\ref{fig:PhaseAreas2}, non-oscillatory).

It is interesting to notice that the discontinuous transition (the boundary between Phases $B$ and $D$) takes place for lower and lower values of the diagonal disorder as the connectivity $c$ decreases, which is due to the fact that the value of $c$ influences the width of the spectral support. In particular, we expect that the discontinuous transition line should tend to $c=1$ in the $d\rightarrow0$ limit, as
the spectral support should exhibit vanishing width at $c=1$. Indeed, for $c<1$ the ensemble consists almost exclusively of a collection of antagonistic trees (with a finite number of unicyclic components)~\cite{janson2011random, dorogovtsev2022nature}. In the absence of diagonal disorder, the former are known to possess a purely imaginary spectrum \cite{valigi2024local}. However, the discontinuous transition shown in the phase diagram in Fig.~\ref{fig:PhaseAreas2} appears to approach zero more rapidly than the $c=1$ limit would suggest. This discrepancy may be attributed to the numerical limitations of the Population Dynamics algorithm for low connectivity regimes discussed at the end of Section~\ref{sec:PopDyn_Algo}.

Among the various features that characterise the shape of the spectral support in the different phases we found, the most important for the complex systems dynamics that inspired our work is probably the nature of the leading eigenvalue, which can either be real or complex. This distinction is relevant both for linear systems (Interaction-like ensemble) and non-linear systems near equilibrium (Jacobian-like ensemble), as it could dictate the dynamical response of the system, which can be either oscillatory or non-oscillatory, depending on whether the imaginary part is non-zero. More generally, since increasing connectivity or diagonal disorder typically expands the spectral support, a stable system whose parameters are changed may eventually reach a point of instability. Depending on the specific spectral phase, the system’s equilibrium will be first destabilised either by monotonic modes or by oscillatory ones. It is important, however, to stress that Jacobian-like matrices might not be representative of the actual Jacobian evaluated at the relevant fixed points of a non-linear system's dynamics. Testing the quality of this approximation and the presence of oscillations in real systems with different connectivities and disorder levels would require extensive numerical experiments, and is left for future work.

Since the possibility of having a complex leading eigenvalue is deeply tied to the presence of the spectral reentrances, it would also be interesting to theoretically investigate their origin, which remains an open question to this day. What seems to happen is that antagonistic interactions in low-connectivity graphs 'repel' the eigenvalues from the region surrounding the real axis, leaving it vacant. It is also interesting to note that, while these reentrances are not observed in random regular graphs\cite{mambuca2022dynamical} (where all nodes have the same degree), they are not exclusive to \ER ensembles, as they also appear in graphs with power-law degree distributions and antagonistic interactions~\cite{valigi2025phdthesis}.

We would like to conclude this section with a comment on the limits that we found to our approach. As discussed at the end of Sections~\ref{sec:cavity_equations} and~\ref{sec:validation}, and in Appendix~\ref{app:Disc}, our estimate for the eigenvalue segment on the real axis suffers from a systematic underestimation; consequently, we employ the support of the diagonal distribution as a proxy to identify the discontinuous transition occurring when the segment overtakes the spectral reentrances. We find similar problems on the whole spectral support when the diagonal disorder is particularly high, but in this case no simple mitigation is available. We argue that these discrepancies are linked to the presence of localised eigenvalues, which are notoriously difficult to capture using Population Dynamics \cite{susca2021cavity, kuhn2008spectra}. More specifically, it has been noted in the literature \cite{metz2019spectral} that investigating the linear stability of the trivial solution via Population Dynamics may fail to correctly identify the singular and pure point components of the spectrum of non-Hermitian random matrices. The development of an algorithm capable of overcoming these limitations, accurately determining both the extent of the segment and the spectral boundary in the high diagonal disorder limit, is currently underway.

\section*{Acknowledgements}

The authors would like to sincerely thank Izaak Neri for valuable insights and ideas that greatly contributed to this work.



\paragraph{Funding information}

P.V. and C.C. acknowledge grants from "Progetti di Ricerca Grandi 2023" (\#RG123188B449C3DE) from Sapienza University of Rome. This project has been supported by the FIS 1 funding scheme (SMaC - Statistical Mechanics and Complexity) from Italian MUR (Ministry of University and Research).


\begin{appendix}
\numberwithin{equation}{section}

\section{Generalised Lotka-Volterra Model} \label{app:GenLotka-Volterra}

The \textbf{generalised Lotka-Volterra model} (GLV) is a model of species-rich ecosystems composed by a pool of $N$ species, each characterised by an abundance $x_i \ge 0$ with $i=1, \dots, N$, interacting through a network of pairwise interactions and evolving according to a set of $N$ coupled dynamical equations:
\begin{equation} \label{eq:GLV}
	\frac{d x_i}{dt}=x_i\left[\left(r_i-\frac{x_i}{K_i}\right)+\sum^N_{j=1;j\neq i}A_{ij}x_j\right] \; .
\end{equation}
The first term is a logistic growth with per capita growth rates $r_i$ and carrying capacities $K_i/r_i$ expressing the self-regulated dynamics of the species in isolation, while the $\{A_{ij}\}$ are the pairwise interactions, which express the effect of the presence of species $i$ on species $j$ and which are gathered in the \emph{interaction matrix} $\matrixbroc{A}$. In general we can identify~\cite{turchin2013complex} five types of pairwise interactions $(A_{ij}, A_{ji})$ among species, depending on the qualitative effect the species $i$ has on $j$ and vice versa, which can be positive $(+)$, negative $(-)$ or null $(0)$: the pair $(\signop(A_{ij}), \signop(A_{ji}))$ can therefore represent predator-prey interaction $(-,+)$, competition $(-, -)$, mutualism $(+, +)$, commensalism $(+, 0)$ and amensalism $(-, 0)$. Accordingly, the types of the interactions set the \emph{sign-pattern} of the interaction matrix. Depending on its structure and sign pattern, determined by the nonzero $A_{ij}$ and their sign, and eventually on the strength of its entries, different ecological models can be obtained and studied, from the completely unstructured ones~\cite{may1972will} to ecosystems in which we provide information on the nature of interactions~\cite{allesina2012stability, allesina2015stability, mambuca2022dynamical, valigi2024local} and up to models with specific structures, like hierarchical food-webs~\cite{poley2023generalized}, cascade models~\cite{poley2024eigenvalue, allesina2015predicting} or multiple islands \cite{patil2024spectral}. For simplicity, and in agreement with the convention on the interactions sign, we can include the intraspecific competition as a negative diagonal entry of the interaction matrix
\begin{equation} \label{eq:GLV_diagonal_entry}
	A_{ii} = - \frac{1}{K_i} \; .
\end{equation}

In real ecosystems performing measurements is, in general, far from trivial. For instance, regarding the interactions between species, their network structure (which species is interacting with which) and their nature (according to the categories defined before) are in some way accessible, while it is much more complicated to quantify their strength~\cite{moore2012energetic, jacquet2016no}. At the same time, when modeling species-rich ecosystems we are generally interested in macroscopic features. In such a scenario, a popular strategy, firstly introduced by Robert May in Ref.~\cite{may1972will}, is to sample randomly these microscopic details, like intraspecific and interspecific interactions, from probability distributions in which we embed the properties we expect to matter.

Various approaches has been developed to investigate the dynamical properties of these ecological models with random parameters~\cite{may1972will, bunin2017ecological, galla2018dynamically, biroli2018marginally, ros2023generalized, roy2019numerical}.

\subsection{Coexistence and different kind of stabilities in ecology} \label{app:stability_ecology}

The possibility to predict and affect the behaviour of ecosystems is of immediate concern for our society, which strongly depends upon them. Therefore the concept of their stability in theoretical ecology has been investigated for decades, leading to the classification of different types of stability of potential practical relevance~\cite{moore2012energetic, dominguez2019unveiling, krumbeck2021fluctuation, grimm1997babel}. In many cases, the stability analysis of a dynamical system can be reduced to a spectral problem involving a matrix~\cite{hahn1967stability}. In this appendix we review several properties relevant to ecological models, specifically feasibility, structural stability, and linear stability. We show how these relate to the spectral properties of the interaction-like $\textbf{A}$ and Jacobian-like $\textbf{J}$ matrices introduced in Section~\ref{sec:Model} of the main text. All these properties are referred to equilibrium points of the dynamics, which are easier to investigate.

\subsection{Feasibility} \label{app:feasibility}

We define \emph{feasibility} as the existence of at least one equilibrium configuration which is ecologically meaningful, \ie, with nonnegative species abundances $x_i \ge 0 \; \forall \; i=1, \ldots, N$ \cite{roberts1974stability, rohr2014structural, grilli2017feasibility, stone2018feasibility, valigi2024local}. Note that generally in a feasible equilibrium several species of the original pool can be extinct, and therefore do not enter in the final composition of the ecosystem. Interestingly, once the extinct species are removed from the system, the search for a fixed point reduces to the linear algebra problem~\cite{valigi2024local}
\begin{equation} \label{eq:glv_equilibrium_surviving}
	\vec{r} = - \matrixbroc{A} \vec{x}^* \; ,
\end{equation}
where $\vec{x}^*$ is the fixed point of the surviving species alone. Note that, although the notation does not make it explicit, the matrix $\matrixbroc{A}$ in Eq.~\eqref{eq:glv_equilibrium_surviving} is restricted to the surviving species and thus generally has a lower dimension than the original interaction matrix in Eq.~\eqref{eq:GLV}.

Therefore, the existence of an equilibrium configuration is granted by the invertibility of the matrix 
$\matrixbroc{A}$, that should be non-singular, \ie, $\det\left(\matrixbroc{A}\right) \neq 0$, which requires that none of its eigenvalue is null\footnote{In the infinitely large $N$ limit, this condition becomes that the continuous part of the spectrum does not include the origin of the complex plane and none of the isolated eigenvalues is null.}. Notably, the spectral reentrances discussed in this paper ensure that the leading eigenvalue does not generally lie on the real axis. This facilitates feasibility even when the matrix spectrum contains eigenvalues with both positive and negative real parts.

Feasibility also requires that all elements in $\vec{x}^*$ are non negative. Naturally, the requirement of having nonnegative abundances is more stringent than the condition for the existence of a solution, yet in the case previously studied the failing of the first condition closely anticipate the breaking of the second~\cite{dougoud2018feasibility}. Following this observation, and in absence of a general rule able to asses full-fledged feasibility, we may consider the condition for the existence of a non trivial $\vec{x}^*$ as a good proxy for feasibility.

\subsection{Ecological Structural Stability} \label{app:structural_stab}

A different kind of stability is represented by ecological \emph{structural stability}~\cite{rohr2014structural, biroli2018marginally, osullivan2019metacommunity, rossberg2017structural, lorenzana2022well} (sometimes also called \emph{press perturbation response}). Broadly speaking, an ecosystem is said to be structurally stable if its dynamical behaviour does not alter with small changes in the model parameters. In fact real ecosystems are expected, in order to persist, to endure small perturbations in environmental conditions without drastically modify their own overall behaviour \cite{grilli2017feasibility}. From a formal point of view the structural stability is related to the volume of parameters space resulting in the same dynamical behaviour \cite{rohr2014structural}. 

An operational approach to investigate this broad property is to look at the stability of the equilibrium abundances of surviving species, $\vec{x}^*$, with respect to small perturbations of the ecological parameters. Also in this case the spectrum of the interaction matrix $\matrixbroc{A}$ is directly relevant. In fact, as shown in Ref.~\cite{valigi2024local}, the susceptibility of $x_i^*$ to little variations $\xi_i$, $\eta_i$ and $\epsilon_{ij}$ of, respectively, the three ecological parameters $r_i$, $K_i$ and $A_{ij}$ is directly related to the inverse of $\matrixbroc{A}$:
\begin{align}
	r_i\rightarrow r_i+\xi_i \quad \Longrightarrow& \quad \frac{\partial x_i^*}{\partial \xi_j}=-(\matrixbroc{A}^{-1})_{ij} \ , \\
	K_i\rightarrow K_i+\eta_i \quad \Longrightarrow& \quad \frac{\partial x_i^*}{\partial \eta_k}\Bigg\rvert_{\vec{\eta}=0}=-(\matrixbroc{A}^{-1})_{ik} \frac{x_k^*}{K_k^2} \ , \\
	A_{ij}\rightarrow A_{ij}+\epsilon_{ij} \quad \Longrightarrow& \quad \frac{\partial x_i^*}{\partial \epsilon_{k\ell}}\Bigg\rvert_{\epsilon=0} = - (\matrixbroc{A}^{-1})_{ik} x_\ell^* \ .
\end{align}
In all the three cases above, a singular behaviour emerges if the spectrum of $\matrixbroc{A}$ contains the origin of the complex plane, hinting to a large susceptibility of the solution of $\vec{x}^*$ to ecological parameters. Again, the reentrance effect discussed in the main text make it possible for an ecological model to remain structurally stable even when certain eigenvalues of the interaction matrix possess positive real parts.

\subsection{Linear Stability}\label{app:linearstability}

Finally, the classical information on \emph{linear stability}, defined as the stability with respect to small perturbations at the level of the abundances (as the ones induced by demographic noise, for instance) which, as discussed in Ref.~\cite{valigi2024local}, is obtained by linearizing the system of dynamical equations around the fixed point $\vec{x}^*$ and hence therein evaluating the {\it Jacobian matrix} $\matrixbroc{J}$, also known in ecology as the {\it community matrix}. In the framework of the generalised Lotka-Volterra model, the Jacobian matrix related to surviving species has elements:
\begin{equation} \label{eq:jacobian_matrix_definition}
	J_{ij}^* = x_i^* A_{ij} \ ,
\end{equation}
with a non trivial stripy structure where the elements in each row are all rescaled by the same factor $x_i^*$. Linear stability requires that the real part of the leading eigenvalue is negative, \ie, $\Re\left(\lambda_1(\matrixbroc{J})\right) < 0$~\cite{hasselblatt2003first, strogatz2024nonlinear}. Moreover, a nonzero imaginary part gives rise to oscillatory dynamics in the vicinity of the equilibrium point with frequency of oscillations proportional to $|\Im[\lambda_1(\matrixbroc{J})]|$. Consequently, the spectral reentrances discussed in this paper may lead to an oscillatory response to perturbations in the case of Jacobian matrices.

\section{The population dynamics algorithm}\label{app:PopDyn}

The population dynamics algorithm is a way of self-consistently solving distributional equations like \eqref{PopDynInt} and \eqref{PopDynJac}  by approximating the target distribution through a large number of data points~\cite{mezard2001bethe}.

In the case of Interaction-like matrices our algorithm is a simple extension of the one found in Ref.~\cite{mambuca2022dynamical}, and reads as follows:
\begin{enumerate}
    \item  start from an initial population of $N_p$ couples $\{(g_j,h_j)\}_{j=1,...,N_p}$ drawn from an arbitrary initial distribution $p_{in}(g,h)$;
    \item draw a residual degree k from $p_{res}(k)$;
    \item uniformly and randomly select $k$ samples $\{(g_i,h_i)\}$ out of the $N_p$ members of the population and generate $k$ i.i.d. couples $\{(u_i,l_i)\}$ following $p_{\alpha}(u_i,l_i)$;
    \item draw a diagonal element $\theta$ from $p_D(\theta)$;
    \item  compute g as follows:
    \begin{gather}\label{gdyn}
        g=-\frac{1}{z-\theta+ \sum_{i=1}^{k}{l_i}g_{i}{u_i}}; 
    \end{gather}
    \item compute h using 
    \begin{equation}\label{pertdyn}
        h=|g|^{2}\displaystyle\sum_{i=1}^{k}h_i |{u_i}|^2;
    \end{equation}
    \item uniformly and randomly select a couple $(g_m,h_m)$ from the available $N_p$ and replace it by (g,h).
\end{enumerate}
In this case we have exploited the fact that the perturbations h and h' are statistically equivalent thanks to the symmetry of $p_\alpha$, thus it is sufficient to study just one of them. We stress the fact that it is fundamental to have a population which is as large as possible so as to obtain a representative sample containing realizations of the pairs $(g,h)$ that are mostly uncorrelated with the other ones. In the rest of this appendix, we will refer to the set of operations (2)-(7) as a Population Dynamics step, and we will call a series of $N_p$ steps a Population Dynamics sweep. Defining the following operations in terms of sweeps rather than steps is more convenient as the former correspond to the number of operations that are necessary - on average - to update the whole population.

As regards the Jacobian-like case, it is very important to preserve the statistical correlations between the abundances $\{x_i\}$ and the other variables: a way to do so is to include the $x_i$-s in the population of the samples, and always updating the quadruples  $(g_m,h_m,h'_m,x_m)$ together and in the correct order. Keeping in mind that the perturbations $h$ and $h'$ are not necessarily equivalent, we modify the algorithm in the following way:
\begin{enumerate}
    \item  start from an initial population of $N_p$ quadruples $\{(g_j,h_j,h'_j,x_j)\}_{j=1,...,N_p}$ drawn from an arbitrary initial distribution\footnote{Since the starting distribution is irrelevant we neglect the correlations between g,h and x at this stage.} $p_{in}(g,h,h'x)=p_{in}(g,h,h')p_X(x)$;
    \item draw a residual degree k from $p_{res}(k)$;
    \item uniformly and randomly select $k$ samples $\{(g_i,h_i,h'_i,x_i)\}$ out of the $N_p$ members of the population and generate $k$ i.i.d. triplets $\{(u_i,l_i,x_i)\}$ following $p(u_i,l_i,x_i)=p_{\alpha}(u_i,l_i)p_X(x_i)$;
    \item draw a diagonal element $\theta$ from $p_D(\theta)$;
    \item draw an abundance $x$ from $p_D(x)$;
    \item  compute g as follows:
    \begin{gather}
        g=-\frac{1}{z+x\theta+ x\sum_{i=1}^{k}{l_i}g_{i}x_i{u_i}}; 
    \end{gather}
    \item compute $h$ and $ h'$ using 
    \begin{equation}\label{pertdynJ}
        h=|g|^{2}\displaystyle\sum_{i=1}^{k}h_i x_i^2|{u_i}|^2 ,\quad h'=|g|^{2}x^2\displaystyle\sum_{i=1}^{k}h'_i |{l_i}|^2\ ;
    \end{equation}
    \item uniformly and randomly select a quadruple $(g_m,h_m,h'_m,x_m)$ from the population and replace it by $(g,h,h',x)$.
\end{enumerate}
In this case, a step refers to operations (2)-(8). 

Once we have a way to correctly approximate the distribution Q, we can determine whether a point z is contained in the spectral support $\mathcal{S}$ or not we employ the following method (for simplicity here we only consider h, but in the Jacobian-like ensemble the study must be carried out for both perturbations):
\begin{enumerate}
    \item after initializing the population, $n_s$ sweeps are performed in order to guarantee the decorrelation from the initial condition;
    \item next, a first average of $|h|$ is performed over $n_r$ sweeps as \begin{equation}
        \overline{|h|}_{1}=\frac{1}{n_rN_p}\sum_{j=1}^{n_r}\sum_{n=1}^{N_p}|h_n^{(j)}|,
    \end{equation}
where the $h_n^{(j)}$-s are the values of h for every element $n$ of the population at each population dynamics sweep $j$;
\item after that, $n_s$ decorrelation sweeps are performed once again;
\item finally, a second average of $|h|$ is measured as before:
\begin{equation}
         \overline{|h|}_{2}=\frac{1}{n_rN_p}\sum_{j=1}^{n_r}\sum_{n=1}^{N_p}|h_n^{(j)}|,
    \end{equation}

\end{enumerate}
After this procedure, if  $\overline{|h|}_{2}>\overline{|h|}_{1}$ then we conclude that $z\in \mathcal{S}$, whereas we infer that  $z\notin \mathcal{S}$ otherwise. Since the dynamics of the perturbation is stochastic, its absolute value can oscillate and only converge or diverge in its average value: for this reason it is important to choose a value of $n_r$ that is large enough to mitigate the effect of such oscillation. Similarly, $n_s$ must be sufficiently high to make the two measurements actually independent of each other and of the initial conditions. Consequently, the choices for $N_p$, $n_s$ and $n_r$ should be tuned taking into account both the necessity of having them as large as possible and the computational cost of the algorithm. Now that we have a way of discerning the points of $ \mathcal{S}$ from the ones outside of it we can find the border of this set by employing the bisection method. In particular, we define the binary value $O_z$ that is equal to 1 if z is outside of $ \mathcal{S}$, 
\begin{equation}\label{outflag}
    O_{z}=\left\{ \begin{aligned} 
  0 \quad \mbox{if} \; z\in \mathcal{S}\\
  1 \quad\mbox{if} \; z\notin  \mathcal{S}
\end{aligned}\right . ,
\end{equation}
and we fix the imaginary part $I$ of the boundary point that we want to find. After that, we perform the following operations:
\begin{enumerate}
    \item fix two extremal points $z_a=(a,I)$ and $z_b=(b,I)$ such that  $O_{z_a}\neq O_{z_b}$;
    \item compute the midpoint $z_c$ and $O_{z_c}$, then replace one of the two extrema with the following rule:
\begin{gather*}
    \mbox{replace } a \mbox{ with } c\mbox{  if  } O_{z_c}= O_{z_a},\\
   \mbox{replace } b \mbox{ with } c\mbox{  if  } O_{z_c}= O_{z_b};
\end{gather*}
\end{enumerate}
We repeat the second operation $N_{iter}$ times and, after that, we choose as our estimate of the border the midpoint between the last extremal points obtained. We have observed that using a fixed $N_{iter}$ rather than allowing for an early stopping whenever $\overline{|h|}_{1}$ and $\overline{|h|}_{2}$ became too similar\footnote{In this sense, a threshold can be either put on $|\overline{|h|}_{1}-\overline{|h|}_{2}|$,  $\big|\log\frac{\overline{|h|}_{1}}{\overline{|h|}_{2}}\big|$  or on other proxies of the difference between the two values.} led to more stable results.  In order to take into account random errors in our estimate of the boundary points, the bisection process is repeated a number $N_{m}\geq10$ of times, and the averages of these realizations, together with the standard error on them, are used in all the computations of the article.

\section{Locating the emergence transition} \label{app:emergence}

In this appendix we focus on locating the emergence transition, which is the is least relevant in terms of the leading eigenvalue's properties. From now on, we will refer to the point in $\mathcal{S}$ with null imaginary part and maximum real part as $z_0$.

First of all, we notice that the observations made for example in the caption of \figref{fig:initial_examples} imply that as $d$ ($d_x$) increases, the segment and the border on the real axis grow at different rates, with the latter being slower than the former as long as $\Re(z_0)>d$. At some point $d^{*,e}$ we have that $\Re(z_0)=d^{*,e}$, which means that the segment has caught up with the border and that it will form the spike for stronger diagonal disorders: the transition point should therefore be defined as the smallest value of $d$ such that the boundary point on the real axis is equal to it.  If we replace $d$ by $d_x$ same reasoning obviously applies for Jacobian-like matrices, with the only difference being that now the distribution of the diagonal elements is centered in $-1$, thus we should be looking for a $d_x^{*,e}$ such that $\Re(z_0)=d_x^{*,e}-1$. Unfortunately, as discussed at the end of Section~\ref{sec:cavity_equations}, the population dynamics algorithm produces an estimate of the boundary points with real part which is a slightly smaller than the expected: this means that our algorithm will yield $\Re(z_0)=d$ (or $\Re(z_0)=d_x-1$) slightly before the segment has actually reached the border of $\mathcal{S}$, hence underestimating the transition point.
\begin{figure}[ht]
    \centering
    \includegraphics[width=0.49\linewidth]{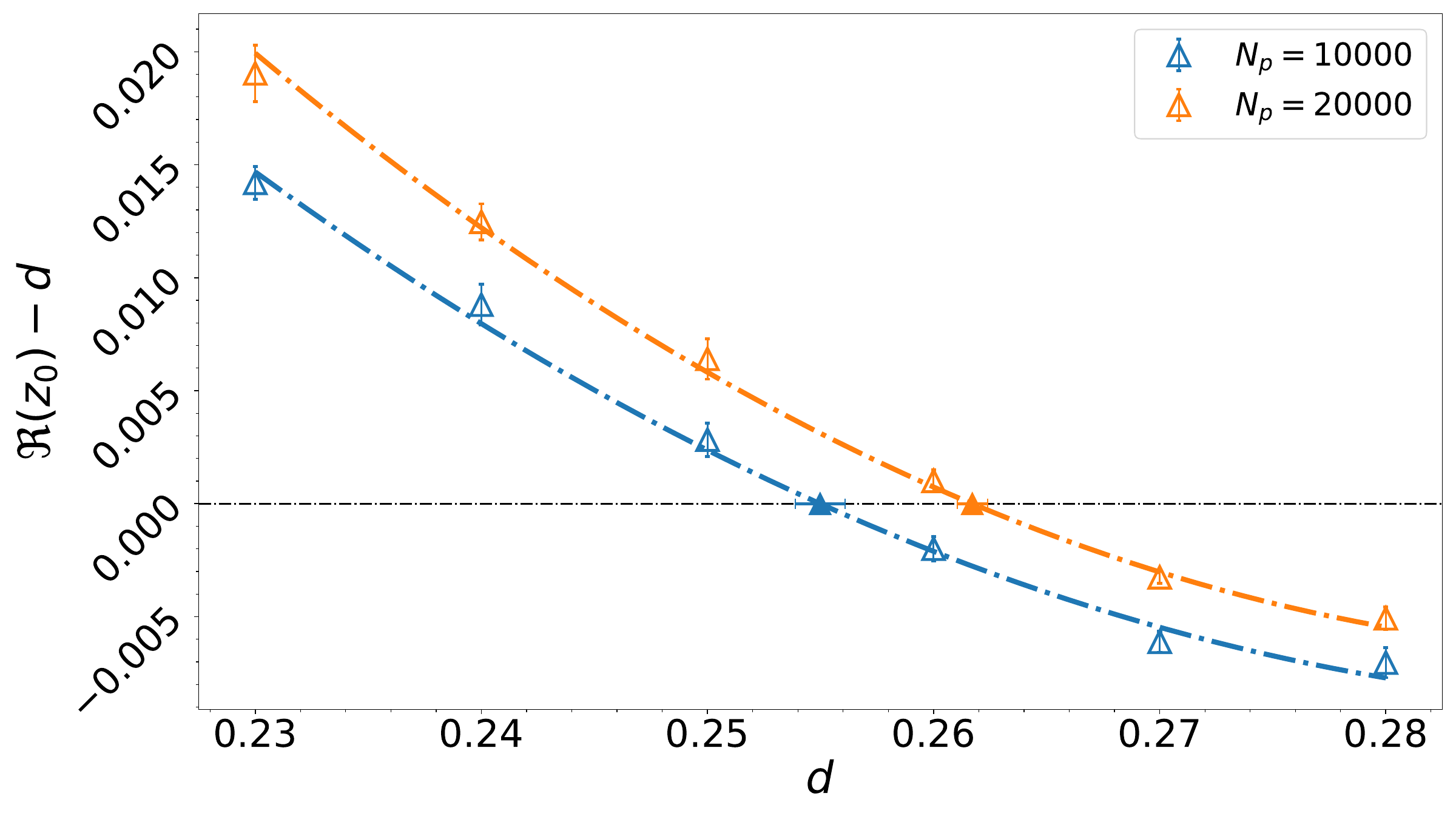}
    \includegraphics[width=0.49\linewidth]{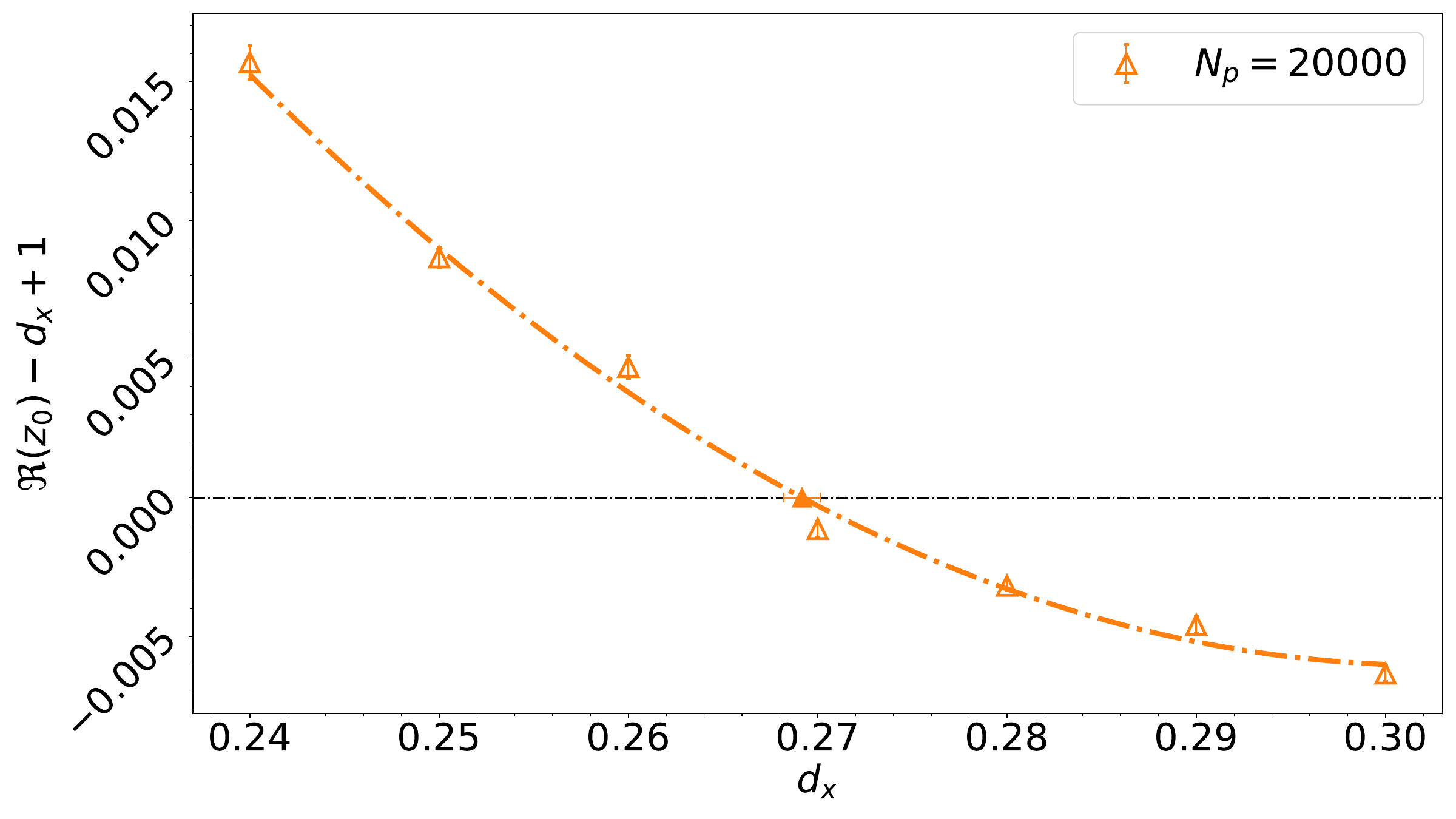}
    \caption{Examples of how we have located the emergence transition point for c=2.2 in both ensembles. The dash dotted lines represent the best fitting curves, while the full markers identify the transition points.}
    \label{fig:LocatingE}
\end{figure}
In \figref{fig:LocatingE} we show the dependence of $\Re(z_0)-d$ (or $\Re(z_0))-d_x+1$) on $d$ ($d_x$): in particular, we observe a monotonically decreasing trend, even after the emergence transition has occurred. This is not would happen in theory, because once the boundary has been reached by the segment their distance should be stuck at 0 for larger values of the disorder, and what we see is caused by the errors that we have already described above. This phenomenon, however, comes in handy for locating the transition point, since we can approximate the trends we have obtained by means of a regular function, and then find the point where it vanishes. This procedure is illustrated in \figref{fig:LocatingE},where we have employed a least squares fit using a quadratic polynomial as fitting function.\\
As it can be seen for the plots related to the interaction matrix\footnote{In this and the following appendices we will always present the results for two different population sizes $(N_P=10000,N_p=20000)$ in the case of Interaction-like matrices and for just one ($N_p=20000$) in the case of Jacobian-like ones. This difference is due to the fact that we have studied the former before the latter: once we have assessed the nature of the finite size effects we have decided to study only one population size in the second case, because of the high computational cost of finding each transition point.}, and as we have already anticipated,  the value of $d^{*,e}$ inferred for $N_p=20000$ is larger than the one obtained for $N_p=10000$, which confirms the presence of finite size effects in the determination of the transition points.

\section{Locating the discontinuous transition} \label{app:Disc}
\begin{figure}[ht]
        \centering
         \begin{subfigure}[b]{0.49\textwidth}
         \centering
         \includegraphics[width=\textwidth]{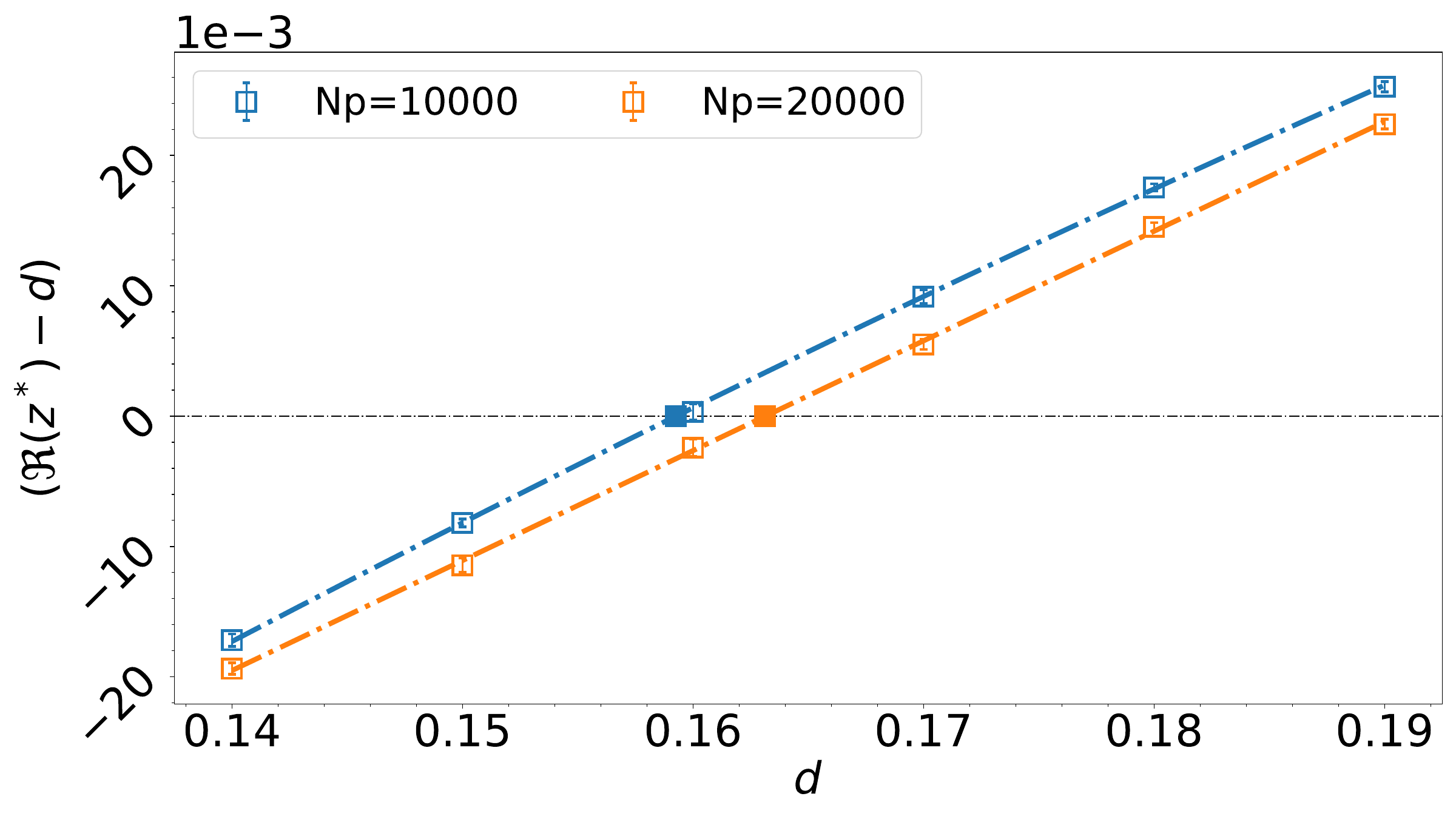}
         \caption{Interaction-like matrices}
         
     \end{subfigure}
     \hfill
     \begin{subfigure}[b]{0.49\textwidth}
         \centering
         \includegraphics[width=\textwidth]{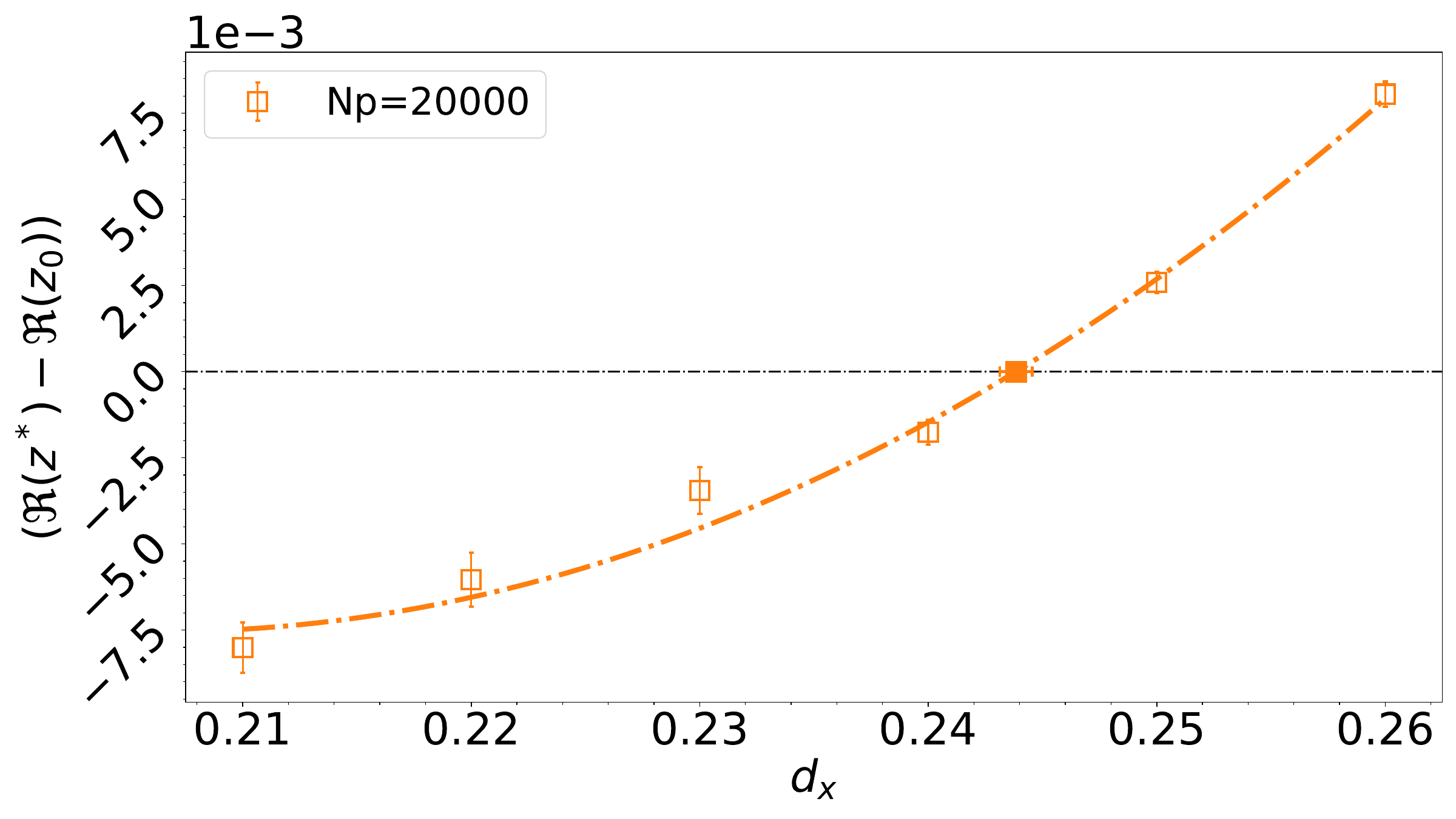}
         \caption{Jacobian-like matrices}
        
     \end{subfigure}
    \caption{($Re(z^*)-\Re(z_0) )\vs d$  (or $d_x$) for the Interaction-like matrices at $c=1.8$ and the Jacobian like ones at $c=2.1$. The dash dotted lines represent the best fitting curves, while the full markers identify the transition points.}
    \label{fig:DiscontinuousExample}
\end{figure}

In this appendix we focus on locating the discontinuous transition of the leading eigenvalue's imaginary part, which takes place when the spike on the real axis becomes wider than the rest of the spectrum, while the reentrance effect is still present. In order to do so, we study the region of the border that does not contain the spike\footnote{This definition is arbitrary, as the spike produces a deformation of the border around it.}, and find the point of maximum real part in it $z^*=\mbox{argmax}({\Re(z)})$: this is our first candidate as leading eigenvalue, to  be compared with the width of the spike. In particular, the discontinuous transition happens at a value $d^{*,d}$  such that $\Re(z^*)=d^{*,d}$, which is also equal to $\Re(z_0)$, as the emergence transition has already occurred. Similarly, the discontinuous transition happens at a value $d_x^{*,d}$ such that $\Re(z^*)=d_x^{*,d}-1=\Re(z_0)$ in the Jacobian-like case.
\begin{figure}[ht]
    \centering
    \includegraphics[width=\linewidth]{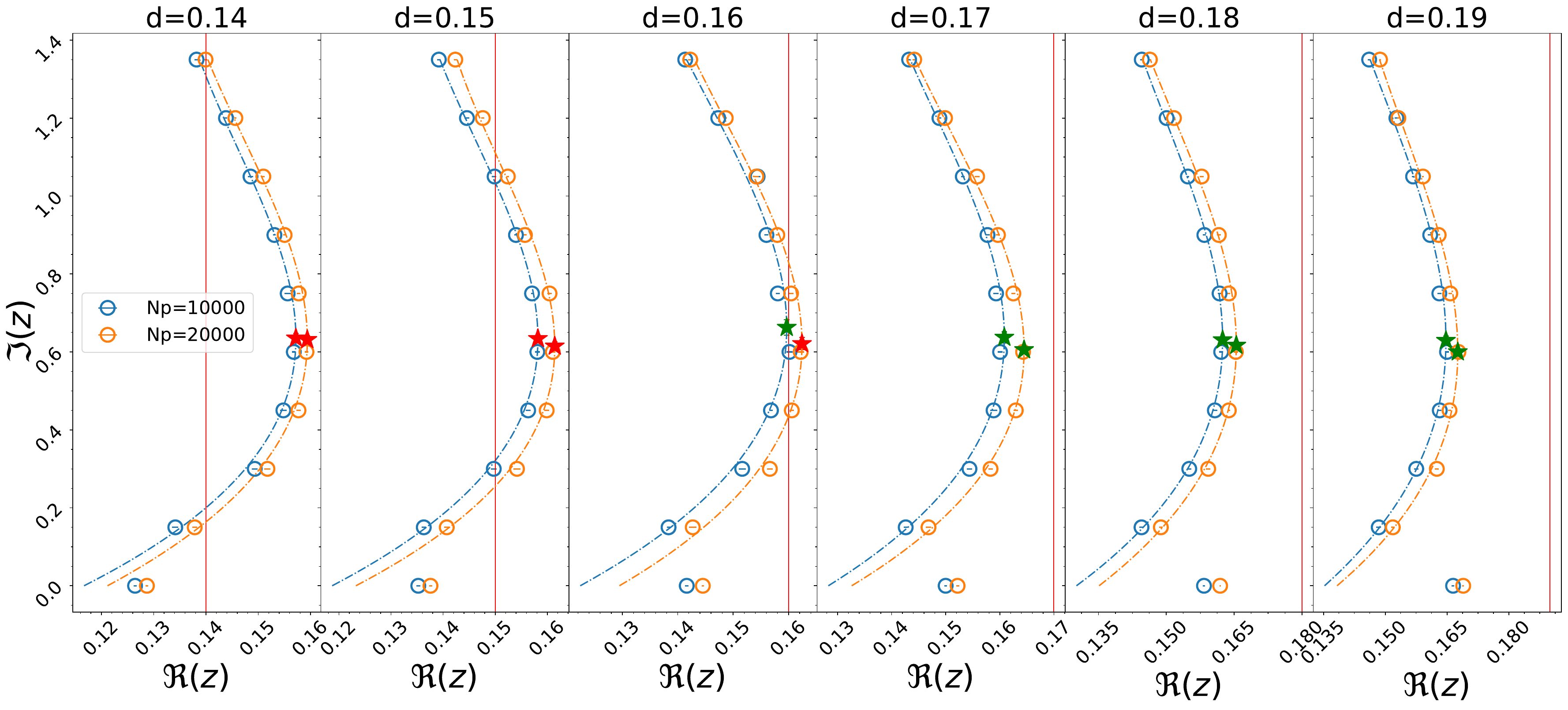}
    \caption{Discontinuous transition at c=1.8 for Interaction-like matrices. The star-shaped markers correspond to $z*$, and we colour them red if the transition has not happened yet, while green otherwise. The vertical lines are placed at $\Re(z_0)$. For both population sizes the results were averaged over $N_m=10$ realisations.}
    \label{fig:DiscontinuousFit}
\end{figure}
\noindent In practice, we first approximate the points of the border away from the real axis by means of a least squares fit, using a third order polynomial as fitting function, \ie \begin{equation}
    \Re(z)=f_d(\Im(z))=a+b\Im(z)+c\Im(z)^22+d\Im(z)^3\,.
\end{equation}
Then, we give our estimate of $z^*$ by finding the point where first derivative of the fitting function vanishes. An example of this procedure is shown in \figref{fig:DiscontinuousFit}, where we only display the interaction case for brevity reasons. After locating $z^*$, we study $(\Re(z^*)-d)\,\vs\, d$ ($(\Re(z^*)-d_x+1\,\vs\, d_x$) and find the point at which it vanishes, as we show in \figref{fig:DiscontinuousExample}. In particular, we employ a least squares fit using a second degree polynomial to fit the data points: this choice has been made to be consistent among all values of $c$, from the smaller ones, at which the curves could be approximated as linear, to the larger ones, at which the curvature of the points is more significant. We stress the fact that, as we have already discussed, different sources of error make us underestimate the real part of the boundary points: this means that we might be underestimating the values of $d^{*,d}$ ($d_x^{*,d}$). We tried mitigating this systematic error by using our estimates of $\Re(z_0)$ instead of $d$ or $d_x-1$, but this only led to a more evident overestimation of the transition points: this can be explained by the fact that estimates of the points on the spike are prone to stronger systematic errors than the ones on the rest of the spectral boundary, as discussed at the end of Sections~\ref{sec:cavity_equations} and~\ref{sec:validation}.
\section{Locating the continuous transition}\label{app:Cont}

\begin{figure}[ht]
    \centering
    \includegraphics[width=.9\linewidth]{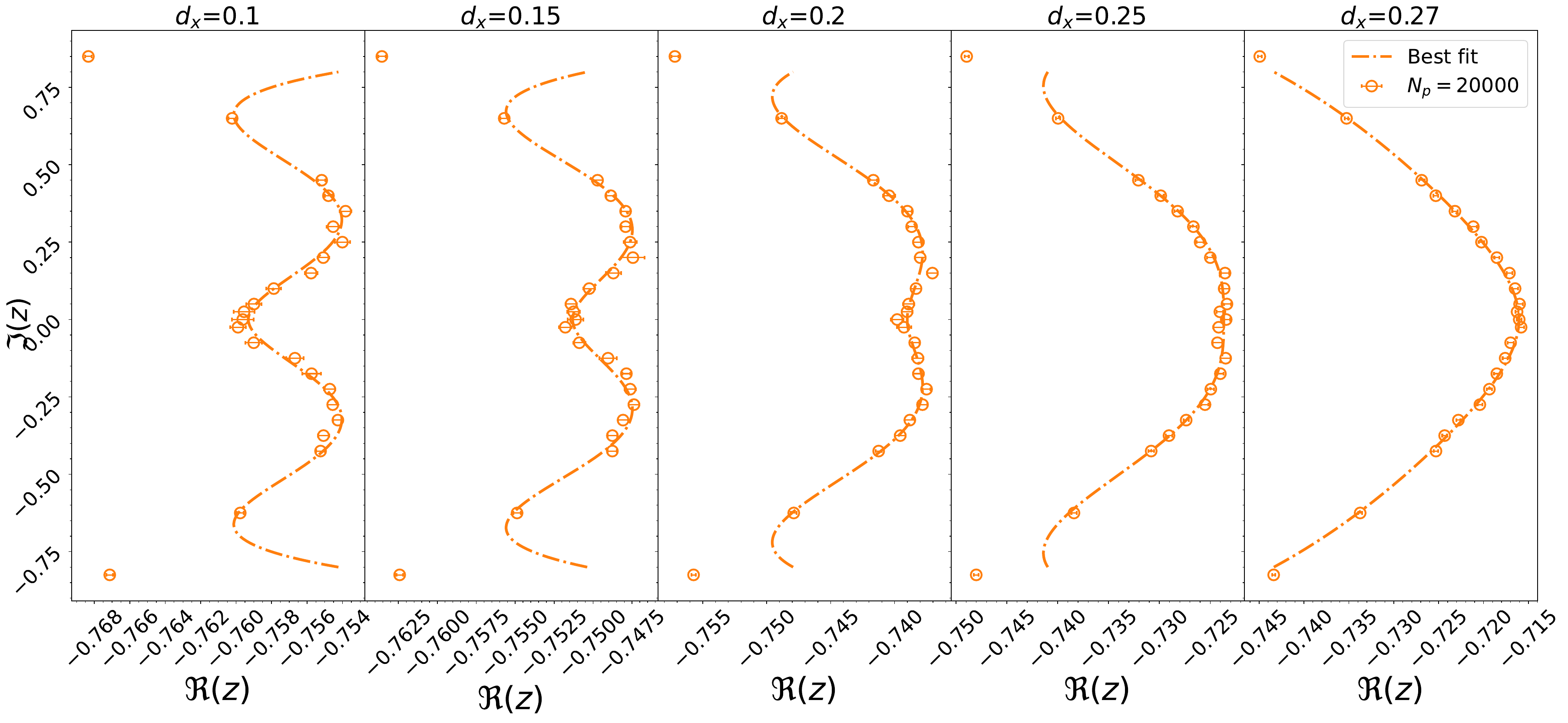}
    \caption{Continuous transition in the Jacobian-like case, for $c=2.3$ and $d_x\in[0.1,0.27]$, for $N_p=20000$. All of the points are averages over $N_m=20$ realisations. In this case the points above and below the real axis have been independently computed, and correspond to slightly different values of $|\Im(z)|$.}
    \label{fig:QuarticFit}
\end{figure}

Locating the continuous transition in $d$ ($d_x$) through the study of the leading eigenvalue is somewhat challenging, as the border becomes flatter and $\lambda^*$ is thus more difficult to locate. It is instead easier to study the concavity of the border near the real axis:  in fact, before the transition happens the indentation is still present and $\mathcal{S}$ is concave, while when the reentrance effect vanishes it becomes convex. In order to do so,  we employ a least squares fit using a quartic polynomial in $|\Im(z)|$ which is differentiable in $z_0$, meaning that the fitting function has the following expression:
\begin{equation}
    \Re(z)=f_c(|\Im(z)|)=a+b|\Im(z)|^2+c|\Im(z)|^3+d|\Im(z)|^4,
\end{equation}
as shown in \figref{fig:QuarticFit}; we then locate the transition point at the value of  $d^{*,c}$ ($d_x^{*,c}$) that makes the second derivative of the polynomial vanish.

 
\begin{figure}[ht]
    \centering
            \centering
         \begin{subfigure}[b]{0.49\textwidth}
         \centering
         \includegraphics[width=\textwidth]{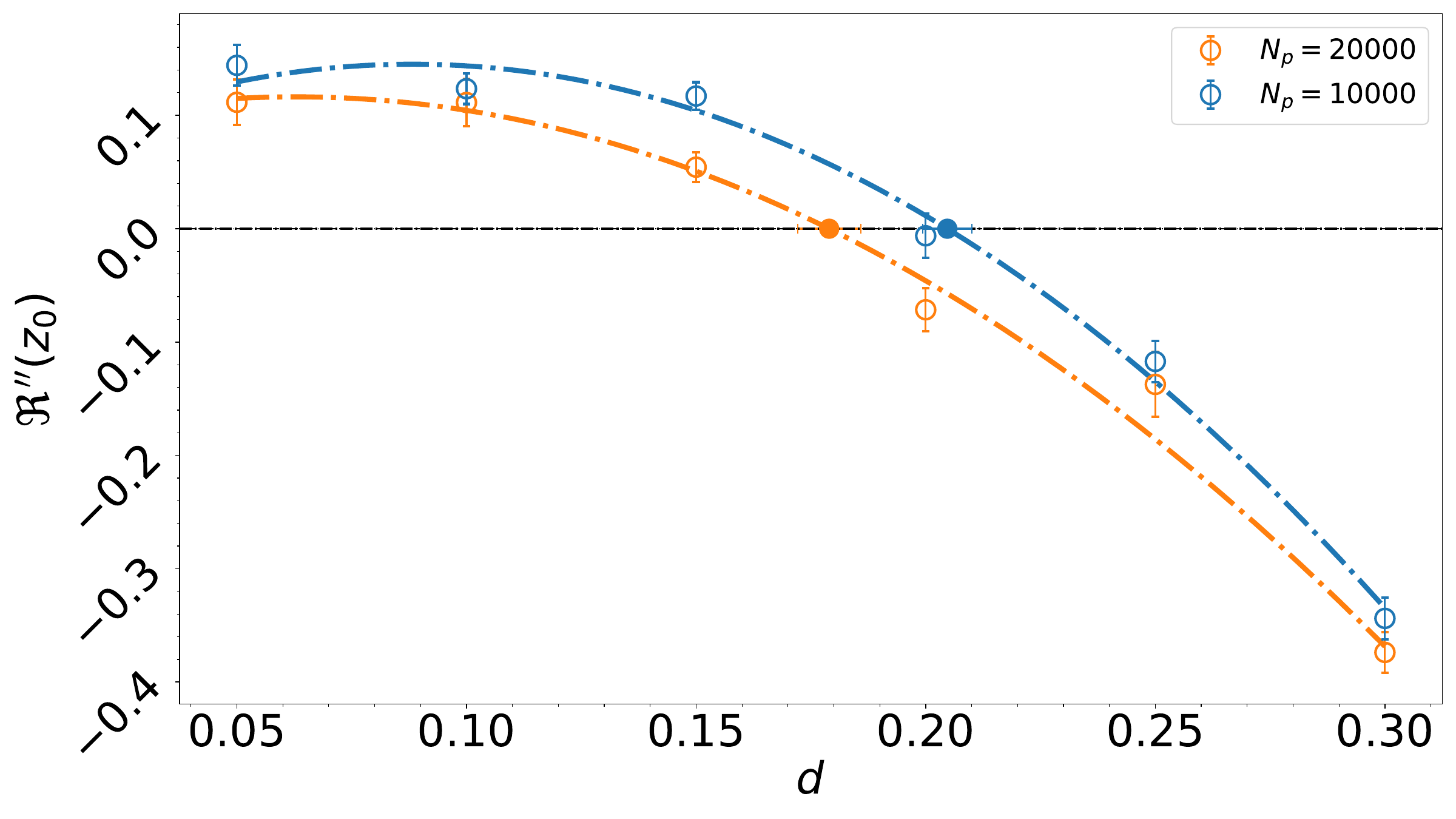}
         \caption{Interaction-like matrices}
         
     \end{subfigure}
     \hfill
     \begin{subfigure}[b]{0.49\textwidth}
         \centering
         \includegraphics[width=\textwidth]{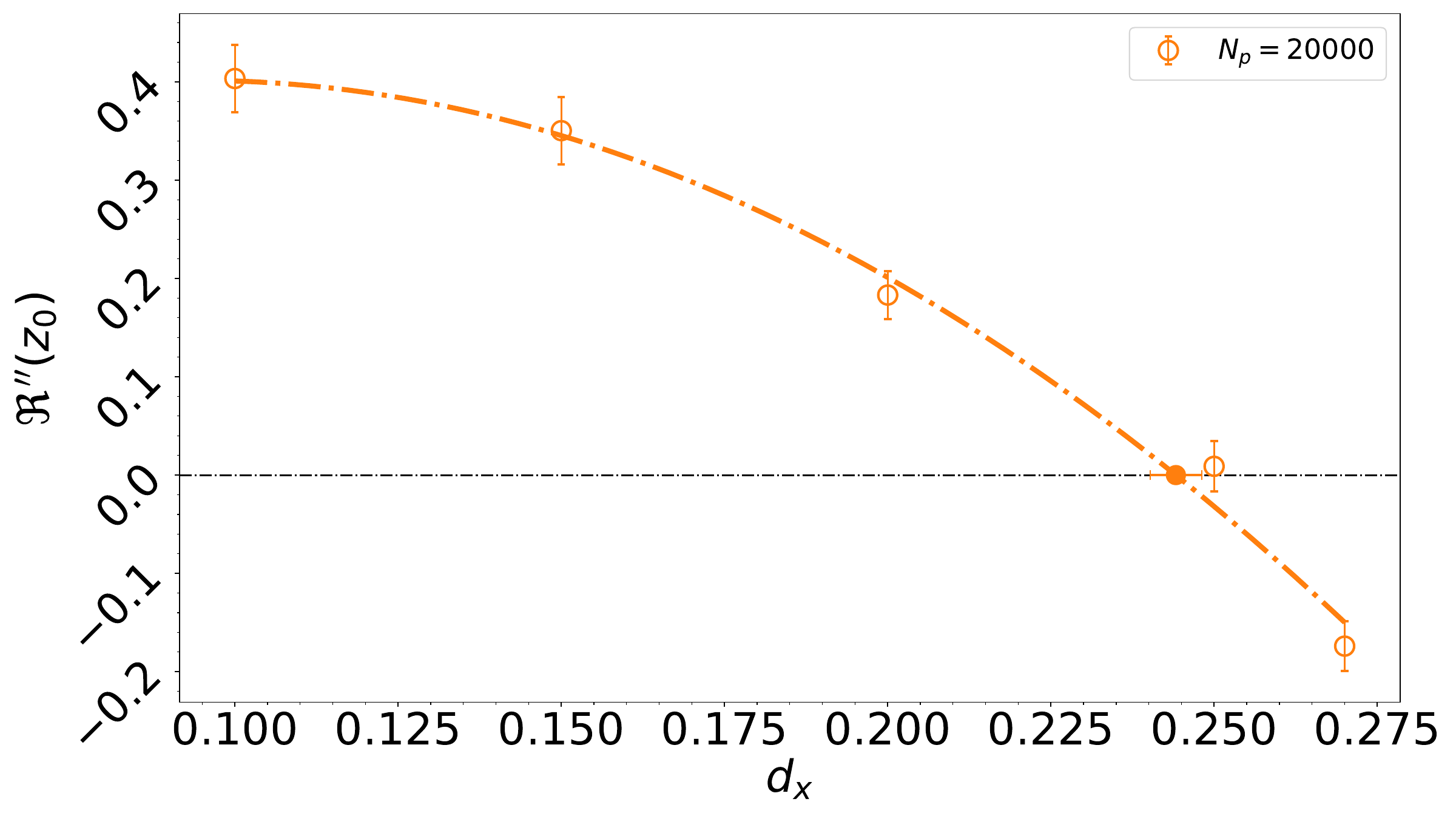}
         \caption{Jacobian-like matrices}
        
     \end{subfigure}
    
    \caption{Continuous transition for the Interaction-like matrices at $c=2.55$, $d\in[0.05,0.3]$ (on the left) and the Jacobian-like ones at $c=2.55$, $d_x\in[0.1,0.275]$ on the right. The dash dotted lines represent the best fitting curves, while the full markers identify the transition points.}
    \label{fig:ContinuousExample}
\end{figure}

 \noindent As it can be seen from the plots of \figref{fig:QuarticFit} the agreement between the best fit curves and the points is not always perfect. Moreover, the comparison between the points at $|\Im(z)|>0.7$ (which were not used in the fit) and the best fit curves highlights the presence of overfitting: this problem, however, appears to mostly affect the region away from the real axis, where the polynomial is dominated by the third and fourth degree terms; on the other hand, the concavity in $z_0$ seems to be reproduced quite well, which is what matters for our study. The approach employed to locate the transition points is analogous to those used in the other sections; specifically, a quadratic polynomial is used to fit the curves $\Re(z)''\vs d$ and $\Re(z)''\vs d_x$, as shown in Figure~\ref{fig:ContinuousExample}. In this case, having observed how sensitive to random errors the procedure was, we have decided to take an average over $N_m= 40$ and $N_m=20$ realizations for the points at $N_p=10000$ and $N_p=20000$ respectively\footnote{In some cases we have actually used more samples than stated, because in our preliminary studies we tried out different values of $N_m$ to optimize this parameter, and then kept the surplus samples in our final analysis.}.
\noindent As we can see from the plot, the finite size effects are more evident in this case than they were in for the discontinuous transition, and act in an opposite way than they did for the emergence one: the transition point at $N_p=10000$ is now located at an higher value of $d$ than the one at $N_p=20000$, which means that we probably overestimating $d^{*,c}$ ($d_x^{*,c}$): this phenomenon is plausibly caused by the fact that, finite size effects are stronger around $\Re(z_0)$, which means that the width of the indentation is overestimated and that the reentrance effect appears to be present even when the border is completely flat. 


\bibliography{bibliography_graphs.bib}

\end{appendix}
\clearpage

\setcounter{equation}{0}
\setcounter{figure}{0}
\setcounter{table}{0}
\setcounter{page}{1}
\setcounter{section}{0}

\renewcommand{\thepage}{S\arabic{page}} 			
\renewcommand{\theequation}{S\arabic{equation}}
\renewcommand{\thefigure}{S\arabic{figure}}
\renewcommand{\thetable}{S\arabic{table}}
\renewcommand{\thesection}{S\arabic{section}}

\input{SI.tex}

\end{document}

%% file: SI.tex
\title{Supplementary Material for: Spectral properties and phase diagrams of sparse antagonistic random matrices with diagonal disorder and Jacobian-like structure}
\author{Luca Giammanco, Pietro Valigi, Chiara Cammarota}

\maketitle
\section{Cavity method and resolvent} \label{SI-supplemental:cavity_and_resolvent}
In this Supplementary Material we will discuss how to obtain closed-form equations that describe the spectrum of sparse, non Hermitian random matrices by using the cavity method under the assumption of tree-likeness. This discussion follows the derivation presented in Ref.~\cite{metz2019spectral}, and we include it to adapt the known results to the notation we use in the main text.

We will at first define the resolvent of a matrix and its relation with the spectrum; next, we will describe the \emph{Hermitization method}, which is a necessary step to study the spectra of non-Hermitian matrices, and the procedure used to derive the equations for the elements of the resolvent. Finally, we will discuss a particular solution of such equations, called the\emph{ trivial solution}, which can be useful for locating the boundary of the spectral distribution's support: in fact, it is only valid and stable outside of this region, thus studying its linear stability allows us to tell whether a point in the complex plane belongs to the spectrum or not.
\subsection{Notation}
 In the following, we will deal with different matrix sizes, mostly $2\times2$, $N\times N$ and $2N\times2N$, and to easily tell apart these cases we introduce an underline to highlight a duplication of the linear size, adopting the following rules:
\begin{itemize}
    \item Scalars, such as $\eta$ or $z$, are denoted by either greek or roman letters;
    \item $2\times2$ blocks, such as $\z$ and $\G_{ij}$, are denoted by underlined roman letters;
    \item $N\times N$ matrices are denoted, as in the main text, by uppercase, boldface letters, like $\textbf{H}$;
    \item $2N\times 2N$ matrices are denoted by uppercase, boldface letters, like $\ub{H}$.
\end{itemize}
\noindent For example, we will start from a matrix $\textbf{H}$ and associate to it a $2N\times 2N$ auxiliary matrix $\ub{H}$, and its permuted version $\tilde{\ub{H}}$; from the latter, we will eventually derive $2\times2$ blocks $\Ht$ having elements $[\Ht]_{\alpha,\beta}$, with $\alpha, \beta=1,2$. \\Finally, lowercase, boldface letters will be used in Schur's formula to denote generic blocks belonging to a bigger matrix, and the identity matrix will always be indicated as a boldface one with its linear size as subscript, like $\textbf{1}_N$.
\subsection{Schur's formula}
The following computations will require the use of Schur's formula for inverting a generic $N\times N$ block matrix \textbf{M} defined as
\begin{equation}
    \textbf{M}=\begin{pmatrix}
        \textbf{a} & \textbf{b}\\
        \textbf{c} &\textbf{d}
    \end{pmatrix},
\end{equation}
where, given two natural numbers $M$ and $L$ such that $N=M+L$, \textbf{b} is an  $M\times L$ block, \textbf{c} is an $L\times M$ one, while \textbf{a} and \textbf{d} are both square matrices of linear size $M$ and $L$ respectively. In order to simplify the final expression for the inverse of \textbf{M}, we now define the its Schur complement $\textbf{s}_\textbf{d}$:
\begin{equation} \label{SI-sd}
    \textbf{s}_\textbf{d}=(\textbf{a}-\textbf{b}\textbf{d}^{-1}\textbf{c})^{-1},
\end{equation}
where $()^{-1}$ denotes the matrix inverse. As it can be confirmed by a direct computation, the inverse of \textbf{M} can now be expressed as
\begin{equation}\label{SI-Schur}
    \textbf{M}^{-1}=\begin{pmatrix}
        \textbf{s}_\textbf{d} && -\textbf{s}_\textbf{d}\textbf{bd}^{-1}\\\\
        -\textbf{d}^{-1}\textbf{cs}_{\textbf{d}} &&\textbf{d}^{-1}+\textbf{d}^{-1}\textbf{cs}_{\textbf{d}}\textbf{b}\textbf{d}^{-1}
    \end{pmatrix}.
\end{equation}
\subsection{Preliminaries on the spectrum of random matrices}

The first step to study the spectrum $\mathscr{S}(\textbf{M})$ of an $N\times N$ random matrix \textbf{M} is to define its empirical spectral distribution $\rho(\textbf{M})$ as
\begin{equation}\label{SI-eq:rho}
    \rho_{\textbf{M}}(z)=\lim_{N\to\infty}\frac{1}{N}\sum_{j=i}^N\delta(z-\lambda_j(\textbf{M})) \, ,\quad z\in\mathbb{C}\, ,
\end{equation}
where $\{\lambda_j(\textbf{M})\}_{j={1\dots N}}$ are the eigenvalues of $\textbf{M}$. If $\rho_{\textbf{M}}(z)$  is a self-averaging quantity, then it will converge to a deterministic probability density function in the infinite size limit, allowing us to make accurate predictions on the behaviour of large systems. For any $N\in\mathbb{N}$, the spectrum of \textbf{M} is composed of a discrete set of eigenvalues, and $\rho(\textbf{M})$ is just a finite sum of Dirac deltas; however, as $N\to \infty$ the spacing of the eigenvalues might go to zero, making \eqref{SI-eq:rho} converge to a probability distribution containing a continuous part. 
When studying the spectra of large matrices it is useful to introduce the \emph{resolvent} or \textbf{Green's function} of a matrix \textbf{M}, defined as\footnote{Sometimes the same quantity is defined with an opposite sign, or the name "Green's function" is used for the trace of the matrix, rather than the matrix itself.}
\begin{equation}\label{SI-Resolvent}
    \textbf{G}_{\textbf{M}}(z)=(\textbf{M}-z\id{N})^{-1},
\end{equation}
which is well-defined only for $z\in\mathbb{C}\setminus\mathscr{S}(\textbf{M})$. It can be shown that the spectral density of \textbf{M} can be inferred from the resolvent by using the following expression\footnote{The proof is based on the fact that $\Tr \textbf{G}_\textbf{M}$ has simple poles at the eigenvalues of \textbf{M}, and it can be found for example in Ref.~\cite{metz2019spectral}.} :
\begin{equation} \label{SI-rhofromG}
    \rho_\textbf{M}(\lambda)=-\frac{1}{\pi N}\pder{}{\overline{z}}\Tr\textbf{G}_{\textbf{M}}(z)\bigg|_{z=\lambda},
\end{equation}
where $\pder{}{\overline{z}}=\big(\der{}{x}+\i\der{}{y}\big)$ is the antiholomorphic derivative, which yields 0 for any analytic function because of the Cauchy-Riemann conditions: as a result, $\rho_\textbf{M}(\lambda)=0\; \forall z\notin\mathscr{S}(\textbf{M})$, which is exactly what we would expect. When studying the spectrum of Hermitian random matrices, which only contains real numbers, it is sufficient to study the resolvent in the proximity of the real axis to avoid its singularities, and obtain a regularised version of the spectral density \cite{metz2019spectral,rogers2010universal}. To be more precise, studying the imaginary part of $\Tr\textbf{G}_{\textbf{M}}(x+\i\tilde{\eta})$, which is well defined $\forall x,\tilde{\eta} \in \mathbb{R}:\tilde{\eta}>0$, gives
\begin{gather}
    \frac{1}{N\pi}\Im\bigg[\Tr\textbf{G}_{\textbf{M}}(x+\i\tilde{\eta})\bigg]=\frac{1}{\pi N}\sum_{i=1}^N\Im\bigg[\dfrac{1}{\lambda(\textbf{M})_{i}-x-\i\tilde{\eta}}\bigg]=\nonumber\\=\frac{\tilde{\eta}}{\pi}\int_{-\infty}^{\infty}\frac{1}{\tilde{\eta}^2+(x-\lambda)^2}\rho_{\textbf{M}}(\lambda)d\lambda\label{SI-regul},
\end{gather}
where we have used the fact that Hermitian matrices are diagonalisable, and we have also exploited the definition of $\rho_\textbf{M}$ provided in Eq. \eqref{SI-eq:rho}.  Since the limit of zero width of a Cauchy-Lorentz distribution is a Dirac delta, taking $\tilde{\eta}\to0$ in Eq. \eqref{SI-regul} provides exactly $\rho_\textbf{M}(x)$. After taking this limit, we can send $N\to\infty$ to obtain the deterministic spectrum; since the regularisation Eq. \eqref{SI-regul} is independent of the size of the matrix, it is safe to exchange the two limits, and it is usually more convenient to deal with the infinite-size resolvent and then retrieve the limiting spectral density.  Unfortunately, the same procedure cannot be applied to non-Hermitian matrices, because the spectrum is not confined to the real axis. An intuitive proof of this fact was given by Sommers et al., who showed that the problem of finding $\Tr\textbf{G}_\textbf{M}$ could be recast as a Poisson equation for an electric field situated outside the region of space containing $\mathscr{S}(\textbf{M})$ and generated by a charge density equal to the spectral one \cite{sommers1988spectrum}. As it is known by Gauss's theorem, however,  the charge distribution does not univocally determine the electric field generated in a region that does not contain any charge.  For these reasons, other ways of tackling the problem are needed in the non Hermitian case. The next section is dedicated to the cavity method, which is particularly effective for sparse, treelike matrices.

\subsection{The cavity method for non-Hermitian Random Matrices}
\subsubsection{An alternative way of expressing the spectral density}
We start with the definition of the $2N\times2N$ auxiliary matrix\footnote{We leave the dependence on $\textbf{M}$ and $\eta$ implicit in order to simplify the notation.} $\ub{H}$:
\begin{equation}\label{SI-auxiliary}
    \ub{H}=\begin{pmatrix}
    (\textbf{M}-z\textbf{1}_N) & -\i\eta\id{N} \\
    -\i\eta\id{N} & (\textbf{M}-z\textbf{1}_N)^\dag \end{pmatrix},
\end{equation}
where $\dag $ denotes the Hermitian conjugate of a matrix and $\eta$ is a small, positive real number that serves as a regulariser and will eventually be put to zero. In particular, $\eta$ is necessary to make $\ub{H}$ invertible $\forall z\in \mathbb{C}$, which is what renders the following calculations well defined. We choose this form of $\ub{H}$ among the many equivalent ones in analogy with the quaternionic approach discussed for example in Refs.~\cite{rogers2010universal} and \cite{gibbs2018effect}, and the $2\times2$ blocks that will be defined in the following can actually be interpreted as quaternions. By applying Eqs. \eqref{SI-sd} and \eqref{SI-Schur}, we can see that the upper-left $N\times N$ block of $\ub{H}^{-1}$ is indeed the resolvent, up to an error quadratic in $\eta$, as we could naively expect by putting $\eta=0$:
\begin{gather}
\textbf{a}=(\textbf{M}-z\textbf{1}_N)\nonumber=\ub{H}_{11},\\
\textbf{b}=\textbf{c}=-\i\eta\id{N}=\ub{H}_{12}=\ub{H}_{21}\nonumber,\\
\textbf{d}=(\textbf{M}-z\textbf{1}_N)^\dag=\ub{H}_{22}\nonumber,\\
    \textbf{s}_\textbf{d}=((\textbf{M}-z\textbf{1}_N)+\eta^2((\textbf{M}-z\textbf{1}_N)^\dag)^{-1})^{-1}=(\textbf{M}-z\textbf{1}_N)^{-1}+O(\eta^2),
\end{gather}
where we have used the known expansion $(\id{N}+\epsilon\textbf{B})^{-1}=\id{N}-\epsilon\textbf{B}+O(\epsilon^2)$, and we have defined the four $N\times N$ blocks of $\ub{H}$ as $\ub{H}_{\alpha\beta}$, with $\alpha,\beta=1,2$. Because of this relation, we can express the empirical spectral density by rewriting \eqref{SI-rhofromG} as
\begin{equation} \label{SI-rhofromH}
    \rho_\textbf{M}(\lambda)=-\frac{1}{\pi N}\lim_{\eta\to 0^+}\bigg\{\pder{}{\overline{z}}\sum_{j=1}^N\big[\ub{H}^{-1}\big]_{jj}\bigg|_{z=\lambda}\bigg\}.
\end{equation}
Once again, we would like to take $N\to\infty$ to obtain the limiting distribution, and it would be easier to do so before sending the regulariser $\eta$ to zero. However, it must be noted that this time the possibility of exchanging the limits is not as obvious as before, and there might be cases when it is not allowed, as noted in Refs.~\cite{rogers2010universal, metz2019spectral}. This issue still constitutes an open line of research. At the same time, the numerical results present in literature show that in many cases it is safe to do so.\\
We will now focus on a technique to find the elements of $\ub{H}^{-1}$ needed for the evaluation of the resolvent. Firstly, we permute the rows and columns of the auxiliary matrix so as to bring close together the elements of $\ub{H}$ that are associated to the same pair of nodes $(j,k)$: to do so, we rearrange the order of the indices by taking the elements $[\textbf{H}_{\alpha\beta}]_{jk}$ from each of the four $N \times N$ blocks of $\ub{H}$, and make them adjacent in the new matrix, as shown below:

\resizebox{0.55\columnwidth}{!}{\parbox{\linewidth}{ \[ \arraycolsep=24pt\def\arraystretch{2.1}
      \left(\begin{array}{ c c c   |  c  c  c}
    \tikzmark{1p1}{[$\ub{H}_{11}]_{11}$}\tikzmark{1p2}{[$\ub{H}_{11}]_{11}$} & \dots &  \tikzmark{1s1}{[$\ub{H}_{11}]_{1N}$}\tikzmark{1s2}{[$\ub{H}_{11}]_{1N}$}\; & \tikzmark{1pHN}{[$\ub{H}_{12}]_{11}$}\tikzmark{2pHN}{[$\ub{H}_{12}]_{11}$} & \dots &  \tikzmark{1sHN}{[$\ub{H}_{12}]_{1N}$}\tikzmark{2sHN}{[$\ub{H}_{12}]_{1N}$}  \\
   \vdots & \ddots & \vdots & \vdots & \ddots & \vdots \\
    \tikzmark{1t1}{[$\ub{H}_{11}]_{N1}$}\tikzmark{1t2}{[$\ub{H}_{11}]_{N1}$} & \dots &  \tikzmark{1q1}{[$\ub{H}_{11}]_{NN}$}\tikzmark{1q2}{[$\ub{H}_{11}]_{NN}$}\; & \tikzmark{1tHN}{[$\ub{H}_{12}]_{N1}$}\tikzmark{2tHN}{[$\ub{H}_{12}]_{N1}$} & \dots &  \tikzmark{1qHN}{[$\ub{H}_{12}]_{NN}$}\tikzmark{2qHN}{[$\ub{H}_{12}]_{NN}$} \\\hline
    \tikzmark{1pVN}{[$\ub{H}_{21}]_{11}$}\tikzmark{2pVN}{[$\ub{H}_{21}]_{11}$} & \dots &  \tikzmark{1sVN}{[$\ub{H}_{21}]_{1N}$}\tikzmark{2sVN}{[$\ub{H}_{21}]_{1N}$}\; & \tikzmark{1pNN}{[$\ub{H}_{22}]_{11}$}\tikzmark{2pNN}{[$\ub{H}_{22}]_{11}$} & \dots &  \tikzmark{1sNN}{[$\ub{H}_{22}]_{1N}$}\tikzmark{2sNN}{[$\ub{H}_{22}]_{1N}$}  \\
   \vdots& \ddots & \vdots & \vdots & \ddots & \vdots \\
    \tikzmark{1tVN}{[$\ub{H}_{21}]_{N1}$}\tikzmark{2tVN}{[$\ub{H}_{21}]_{N1}$} & \dots &  \tikzmark{1qVN}{[$\ub{H}_{21}]_{NN}$}\tikzmark{2qVN}{[$\ub{H}_{21}]_{NN}$}\; & \tikzmark{1tNN}{[$\ub{H}_{22}]_{N1}$}\tikzmark{2tNN}{[$\ub{H}_{22}]_{N1}$} & \dots &  \tikzmark{1qNN}{[$\ub{H}_{22}]_{NN}$}\tikzmark{2qNN}{[$\ub{H}_{22}]_{NN}$}
  \end{array}\right)
  \Highlighta{1p1}{1p2} \Highlightb{1s1}{1s2} \Highlightc{1t1}{1t2} \Highlightd{1q1}{1q2}
  \Highlighta{1pHN}{2pHN} \Highlightb{1sHN}{2sHN} \Highlightc{1tHN}{2tHN} \Highlightd{1qHN}{2qHN}
  \Highlighta{1pVN}{2pVN} \Highlightb{1sVN}{2sVN} \Highlightc{1tVN}{2tVN} \Highlightd{1qVN}{2qVN}
  \Highlighta{1pNN}{2pNN} \Highlightb{1sNN}{2sNN} \Highlightc{1tNN}{2tNN} \Highlightd{1qNN}{2qNN}
  \;
  \smash{
\begin{tikzpicture}
    \node[single arrow, draw=red, fill=red, 
       minimum width = 3pt, single arrow head extend=3pt,
      minimum height=15mm] {}; 
  \end{tikzpicture}}\;
  \arraycolsep=7pt\def\arraystretch{2}
  \left(\begin{array}{c c c c c c}
    \tikzmark{1p1}{$[\ub{H}_{11}]_{11}$} & [\ub{H}_{12}]_{11} & \dots &  \dots &\tikzmark{1s1}{$[\ub{H}_{11}]_{1N}$}  &  [\ub{H}_{12}]_{1N} \\\;
     [\ub{H}_{21}]_{11} &\tikzmark{1p2}{$[\ub{H}_{22}]_{11}$} & \dots &  \dots   &  [\ub{H}_{21}]_{1N}&\tikzmark{1s2}{$[\ub{H}_{22}]_{1N}$} \\
    \vdots & \vdots &  \ddots & \ddots  & \vdots &  \vdots \\
    \vdots & \vdots &  \ddots & \ddots & \vdots &  \vdots  \\
   \tikzmark{1t1}{$[\ub{H}_{11}]_{N1}$} & [\ub{H}_{12}]_{N1} & \dots & \dots & \tikzmark{1q1}{[$\ub{H}_{11}]_{NN}$} & \ub{H}_{12}]_{NN} \\\;
    [\ub{H}_{21}]_{N1} & \tikzmark{1t2}{$[\ub{H}_{22}]_{N1}$} &  \dots & \dots & \ub{H}_{21}]_{NN} &  \tikzmark{1q2}{[$\ub{H}_{22}]_{NN}$}
  \end{array}\right)
  \Highlighta{1p1}{1p2} \Highlightb{1s1}{1s2} \Highlightc{1t1}{1t2} \Highlightd{1q1}{1q2}
\]}}\\
In the new matrix thus obtained, which we call $\tilde{\ub{H}}$,  the elements of  $\ub{H}$ that were once separated by $N$ rows and/or columns due to the enlarged size of $\ub{H}$ can now be grouped in $2\times2$ blocks that we define in the following way:
\begin{gather}
    \tilde{\ub{H}}=\begin{pmatrix}
\Ht_{11} & \dots & \Ht_{1j}  & \dots & \Ht_{1N}\\
\vdots    & \ddots&  &  & \vdots\\
\Ht_{j1} &  & \Ht_{jj} &  &\Ht_{jN} \\
\vdots  & &   & \ddots & \vdots\\
\Ht_{N1}  & \dots & \Ht_{Nj} & \dots&\Ht_{NN} \\
    \end{pmatrix},\\
        \Ht_{jk}=\begin{pmatrix}
        [\tilde{\ub{H}}]_{2j-1,2k-1} & [\tilde{\ub{H}}]_{2j-1,2k}\\ [\tilde{\ub{H}}]_{2j,2k-1} 
         & [\tilde{\ub{H}}]_{2j,2k}
    \end{pmatrix}=\begin{pmatrix}
        [\ub{H}]_{j,k} & [\ub{H}]_{j,k+N}\\ [\ub{H}]_{j+N,j} 
         & [\ub{H}]_{j+N,k+N}
    \end{pmatrix}=\nonumber\\=\begin{pmatrix}
        [\textbf{H}_{11}]_{j,k} & [\textbf{H}_{12}]_{j,k}\\ [\textbf{H}_{21}]_{j,k} 
         & [\textbf{H}_{22}]_{j,k}
    \end{pmatrix}=\M_{jk}-\z\delta_{jk}.\label{SI-HDef}
\end{gather}
Where we have introduced the $2\times 2$ matrices $\{\M_{jk}\}$ and $\z$,
\begin{equation}\label{SI-Mzdef}
    \M_{jk}= \begin{pmatrix}
    [\textbf{M}]_{jk} & 0 \\
    0 & \overline{[\textbf{M}]_{kj}} \end{pmatrix}
     \quad , \quad  \z= \begin{pmatrix}
    z & i\eta \\
    i\eta & \overline{z} \end{pmatrix}.
\end{equation}
We now move on to the evaluation of $\tilde{\ub{H}}^{-1}$, which, just like its inverse, can be divided into $N$ blocks of dimensions $2\times 2$. In order to express them, we notice that the permutation operation described above is just a relabelling of the basis vectors based on the following rule:
\begin{equation}\label{SI-perm}
      (j+(\alpha-1)N,k+(\beta-1)N)\rightarrow (2j+\alpha-2 ,2k+\beta-2), \; \forall j,k =1,...,N \; \forall \alpha, \beta =1,2 \,,
\end{equation}
 which commutes with the matrix inversion. This fact can be easily seen by representing the permutation as an orthogonal matrix $\underline{\textbf{P}}$, and exploiting the basic rules of matrix algebra to write
\begin{equation}
\tilde{\ub{H}}=\underline{\textbf{P}}\,\ub{H}\underline{\textbf{P}}^T\implies\tilde{\ub{H}}^{-1}=\underline{\textbf{P}}\,\ub{H}^{-1}\underline{\textbf{P}}^T,
\end{equation}
where by $\underline{\textbf{P}}^T$ we denote the transpose of $\underline{\textbf{P}}$. By exploiting this property we can have a straightforward expression for the blocks of $\tilde{\ub{H}}^{-1}$, which we call $\{\G_{jk}\}$:
\begin{equation}
    \tilde{\ub{H}}^{-1}=\begin{pmatrix}
\G_{11} & \dots & \G_{1j}  & \dots & \G_{1N}\\
\vdots    & \ddots&  &  & \vdots\\
\G_{j1} &  & \G_{jj} &  &\G_{jN} \\
\vdots  & &   & \ddots & \vdots\\
\G_{N1}  & \dots & \G_{Nj} & \dots&\G_{NN} \\
    \end{pmatrix},\\\end{equation} 
    \begin{equation}
    \label{SI-Gblocks}
        \G_{jk}=\begin{pmatrix}
        [\tilde{\ub{H}}^{-1}]_{2j-1,2k-1} & [\tilde{\ub{H}}^{-1}]_{2j-1,2k}\\ & \\ [\tilde{\ub{H}}^{-1}]_{2j,2k-1} 
         & [\tilde{\ub{H}}^{-1}]_{2j,2k}
    \end{pmatrix}=\begin{pmatrix}
        [\ub{H}^{-1}]_{j,k} & [\ub{H}^{-1}]_{j,k+N}\\ & \\ [\ub{H}^{-1}]_{j+N,j} 
         & [\ub{H}^{-1}]_{j+N,k+N}
    \end{pmatrix} \, .
\end{equation}
From the previous expression we immediately see that $[\ub{H}^{-1}]_{jj}=[\G_{jj}]_{11}$ and we can thus rewrite \eqref{SI-rhofromH} as
\begin{equation} \label{SI-rhofromGB}
    \rho_\textbf{M}(\lambda)=-\frac{1}{\pi N}\lim_{\eta\to 0^+}\bigg\{\pder{}{\overline{z}}\sum_{i=1}^N\big[\G_{jj}\big]_{11}\bigg|_{z=\lambda}\bigg\}.
\end{equation}
The last step to determining the spectral density is therefore computing the diagonal blocks of $\tilde{\ub{H}}^{-1}$, a problem that can be tackled by iteratively applying Schur's formula.
\subsubsection{Equations for inverting the auxiliary matrix}
In order to employ Schur's formula to our advantage we at first rearrange once again the rows and columns of $\ub{H}^{-1}$, keeping in mind that this operation commutes with the matrix inversion: our goal here is to isolate the elements related to a chosen node $j$ from the rest of the matrix, thus obtaining
\begin{gather}
\begin{pmatrix}
    \Ht_{jj} & \Ht_{j1} &\dots & \Ht_{j,j-1} & \Ht_{j,j+1}& \dots&\Ht_{j,N} \\
    \Ht_{1j} &          &          &          &          &          &           \\
    \vdots &          &          &          &          &          &           \\
    \Ht_{j-1,j} &          &          & \tilde{\ub{H}}^{(j)}&          &          &           \\
    \Ht_{j+1,j} &          &          &          &          &          &           \\
    \vdots &          &          &           &          &          &           \\
    \Ht_{N,j} &          &          &           &          &          &           \\
\end{pmatrix}^{-1}=\label{SI-HBlocks}\\
=
\begin{pmatrix}
    \G_{jj} & \G_{j1} &\dots & \G_{j,j-1} & \G_{j,j+1}& \dots&\G_{j,N} \\
    \G_{1j} &     \G_{11}     &          &          &          &          &           \\
    \vdots &          &     \ddots     &          &          &          &    \vdots       \\
    \G_{j-1,j} &          &          &\G_{j-1,j-1}&          &          &    \G_{j-1,N}       \\
    \G_{j+1,j} &          &          &          & \G_{j+1,j+1}         &          &   \G_{j+1,N}        \\
    \vdots &          &          &           &          &   \ddots       &  \vdots         \\
    \G_{N,j} &   \G_{N1}      &     \dots     &   \G_{N,j-1}        &   \G_{N,j+1}        &     \dots     &    \G_{N,N}       \\
\end{pmatrix},
\end{gather}
where $\tilde{\ub{H}}^{(j)}$ is the $2(N-1)\times 2(N-1)$ matrix obtained after removing the (2j-1)-th and 2j-th rows and columns from  $\tilde{\ub{H}}$, which is equivalent to removing the node j from the interaction network: this procedure leaves a "cavity" in the original graph $\mathscr{G}$, and yields a new one that is called for this reason the \textbf{cavity graph}, $\mathscr{G}^{(j)}$. 
In order to apply Schur's formula and retrieve the equations for $\{\G_{jk}\}$, we divide \eqref{SI-HBlocks}in the following blocks:
\begin{gather}
\textbf{a}=\Ht_{jj}\nonumber,\displaybreak[0]\\
\textbf{b}=\begin{pmatrix}
    \Ht_{j1}\;, &\dots\;  & \Ht_{j,j-1}\;,  & \Ht_{j,j+1} \;, & \dots&\Ht_{j,N}
\end{pmatrix}\nonumber ,\displaybreak[0]\\
\textbf{c}=\begin{pmatrix}
    \Ht_{1j}\;,  &\dots & \Ht_{j-1,j}\;,  & \Ht_{j+1,j}\;, & \dots&\Ht_{N,j}
\end{pmatrix}^{T}\nonumber,\\
\textbf{d}=\tilde{\ub{H}}^{(j)}\nonumber.
\end{gather}
From these we get
\begin{gather}
    \G_{jj}=\textbf{s}_\textbf{d}=(\textbf{a}-\textbf{b}\textbf{d}^{-1}\textbf{c})^{-1},\label{SI-DiagGd} \\
    \begin{pmatrix}
    \G_{j1}\;, &\dots\;  & \G_{j,j-1}\;,  & \G_{j,j+1} \;, & \dots&\G_{j,N}
\end{pmatrix}=-\textbf{s}_\textbf{d}\textbf{b}\textbf{d}^{-1}.\label{SI-OffDiagGd}
\end{gather}
Both equations require a more straightforward way to deal with the matrix inverse of $\tilde{\ub{H}}^{(j)}$, which we can divide into $N-1$ blocks called $\{\G^{(j)}_{kl}\}$. Since we have removed the j-th node we do not use this index to label the rows and columns of the new matrix: this way, we keep the connection between the labels of the blocks and the nodes they refer to straightforward. By doing so, we get the following expression:
\begin{gather}\label{SI-CavityG}
\bigg(\tilde{\ub{H}}^{(j)}\bigg)^{-1}=
\begin{pmatrix}
    \G^{(j)}_{11} & \dots & \G^{(j)}_{1,j-1} & \G^{(j)}_{1,j+1}& \dots&\G^{(j)}_{1,N} \\
    \vdots &    \ddots     &          &          &          &    \vdots       \\
    \G^{(j)}_{j-1,1} &         &\G^{(j)}_{j-1,j-1}&          &          &    \G^{(j)}_{j-1,N}       \\\\
    \G^{(j)}_{j+1,1} &         &          & \G^{(j)}_{j+1,j+1}         &          &   \G^{(j)}_{j+1,N}        \\
    \vdots &          &           &          &   \ddots       &  \vdots         \\
    \G^{(j)}_{N,1} &     \dots     &   \G^{(j)}_{N,j-1}        &   \G^{(j)}_{N,j+1}        &     \dots     &  \G^{(j)}_{NN}   \\
\end{pmatrix}.
\end{gather}
Note that, in general, $\G^{(j)}_{jk}\neq \G_{jk}$, as they are obtained by inverting two different matrices. By inserting the last expression into \eqref{SI-DiagGd} and \eqref{SI-OffDiagGd} we get the following equations, both of which are valid for any kind of matrix:
\begin{gather}
    \G_{jj}=-\bigg(\z-\M_{jj}+ \sum_{\substack{k=1\\k\neq j}}^N\sum_{\substack{i=1\\i\neq j}}^N\M_{jk}\G^{(j)}_{ki}\M_{ij}\bigg)^{-1},\label{SI-diagG}\\
    \G_{jk}=-\G_{jj}\sum_{i=1,i\neq{j}}^N\M_{ji}\G^{(j)}_{ik}, \label{SI-offdiagG}
\end{gather}
where we have used Eqs. \eqref{SI-HDef}, \eqref{SI-Mzdef}, and standard manipulations to express the matrix products. After removing a node $j$, the problem of determining the elements of \eqref{SI-CavityG} on the cavity graph is formally equivalent to the one we have just discussed, thus we can easily write the following equations:
\begin{gather}
    \G^{(l)}_{jj}=-\bigg(\z-\M_{jj}+ \sum_{\substack{k=1\\k\neq j,l}}^N\sum_{\substack{i=1\\i\neq j,l}}^N\M_{jk}\G^{(l,j)}_{ki}\M_{ij}\bigg)^{-1}, \label{SI-diaGGCav}\\
    \G^{(l)}_{jk}=-\G_{jj}^{(l)}\sum_{\substack{i=1\\i\neq{j,l}}}^N\M_{ji}^{(l)}\G^{(l,j)}_{ik}, \label{SI-offdiagGcav}
\end{gather}
where the $\{\G^{(l,j)}_{ik}\}$ are the blocks of the inverse of the (permuted) auxiliary matrix obtained after removing the $i$-th and $j$-th nodes from the interaction network $\mathscr{G}$. As it can be easily inferred, the equations we have obtained exist in a hierarchy, meaning that the blocks related to a graph with $n$ cavities depend on the ones obtained by creating $n+1$ cavities. In order to break this hierarchy we need to better understand the relations between the blocks we are dealing with and the topology of the underlying interaction network, and then to exploit the tree-likeness of our model.
\subsubsection{Cavity equations for tree-like matrices}
In this section we will focus on tree graphs and exploit their properties to delve deeper into the connections between the interaction network and the values of the matrices $\{\G_{jk}\}$, $\{\M_{jk}\}$ and their "cavity counterparts". After that, we will justify why the results we obtain can be also used for networks that are only locally tree-like, as the \ER ones.\\  First of all, we note that \eqref{SI-HDef} implies that $\M_{jk}$ is different from zero only if the two nodes are connected, \ie:
\begin{equation}\label{SI-condM}
    \M_{jk}\neq0 \iff k\in \partial_j \iff [\textbf{C}]_{jk}=[\textbf{C}]_{kj}=1.
\end{equation}
Furthermore, if we plug \eqref{SI-offdiagGcav} into \eqref{SI-offdiagG} we obtain the following expression:
\begin{gather}
\G_{jk}=\G_{jj}\bigg[-\M_{jk}\G^{(j)}_{kk}+\sum_{\substack{i=1\\i\neq{j,k}}}^N\M_{ji}\G_{ii}^{(j)}\M_{ik}^{(j)}\G^{(j,i)}_{kk}+ \sum_{\substack{i,i'=1\\i\neq{j,k}\\i'\neq i,j,k}}^N\M_{ji}\G_{ii}^{(j)}\M_{ii'}^{(j)}\G^{(j,i)}_{i'k}\bigg] \, .
\end{gather}  
Because of \eqref{SI-condM}, the first term of the last formula is different from zero if and only if there is an edge between j and k, whereas the second represents the contribution of all the existing paths of length 2 between the nodes (going back to the same vertex is not possible because of the cavities). The third term can be expanded in a similar way, by adding another cavity to \eqref{SI-offdiagGcav} and getting an expression for $\G^{(j,i)}_{i'k}$:
\begin{gather}
    \sum_{\substack{i,i'=1\\i\neq{j,k}\\i'\neq i,j,k}}^N\M_{ji}\G_{ii}^{(j)}\M_{ii'}^{(j)}\G^{(j,i)}_{i'k}=\nonumber\\=\sum_{\substack{i,i'=1\\i\neq{j,k}\\i'\neq i,j,k}}^N\M_{ji}\G_{ii}^{(j)}\M_{ii'}^{(j)}\G^{(j,i)}_{i'i'}\M_{i'i''}^{(j,i)}\G^{(j,i,i')}_{kk}+\sum_{\substack{i,i',i''=1\\i\neq{j,k}\\i'\neq i,j,k\\i''\neq{i',i,j,k}}}^N\M_{ji}\G_{ii}^{(j)}\M_{ii'}^{(j)}\G^{(j,i)}_{i'i'}\M_{i'i''}^{(j,i)}\G^{(j,i,i')}_{i''k} \; ;
\end{gather}
once again, the first term corresponds to the contributions given by all possible paths of length 3 between $j$ and $k$, while the second must be expanded further. By iterating this procedure we can see that the off-diagonal blocks $\{\G_{jk}\}$ are computed by summing terms corresponding to the existing paths of all possible legths linking $ j$ and $k$: on the other hand, if there is no such path the considered block is zero, hence we can conclude that 
\begin{equation}\label{SI-condG}
    \G_{jk}\neq 0 \implies \mbox{j and k belong to the same connected component in $\mathscr{G}$}.
\end{equation}
The previous results hold for the cavity blocks too if we consider the cavity graph instead of the original one. Taking the conditions \eqref{SI-condM} and \eqref{SI-condG} for the cavity elements $\G_{kk}^{(j)}$ and $\G_{ki}^{(j)}$, and putting them together, we can immediately derive the following implications:
\begin{gather}
    \M_{jk}\G^{(j)}_{kk}\M_{kj}\neq0 \implies k\in \partial_j,
\\
\M_{jk}\G^{(j)}_{ki}\M_{ij}\neq0 \implies \exists\mbox{ a cycle in  } \mathscr{G} \mbox{  of the form  } j\to k \to \{n\} \to i\to j,
\end{gather}
where $\{n\}$ denotes any possible path of length larger than one not containing i, j or k. Since trees do not contain cycles by definition, the second condition is never true in the graphs we are now considering, thus equations \eqref{SI-diagG} and  \eqref{SI-diaGGCav} can be further simplified:
\begin{gather}
    \G_{jj}=-\bigg(\z-\M_{jj}+ \sum_{k\in\partial_j}\M_{jk}\G^{(j)}_{kk}\M_{kj}\bigg)^{-1},\label{SI-diagGtrees}\\
    \G^{(l)}_{jj}=-\bigg(\z-\M_{jj}+ \sum_{k\in\partial_j	\setminus\{l\}}\M_{jk}\G^{(l,j)}_{kk}\M_{kj}\bigg)^{-1}\label{SI-diaGGCavtrees}.
\end{gather}
The final step to break the hierarchy and get a set of closed-form equations is to perform the following substitution:
\begin{equation}\label{SI-BreakingHier}
    \G^{(l,j)}_{kk}=\G^{(j)}_{kk} \quad \mbox{if  } k\in\partial_j\setminus\{l\}.
\end{equation}
This is motivated by the fact that when j is removed and a cavity is formed, the connected component containing the node is divided in $k_j$ mutually disconnected subgraphs, each of them containing one former neighbour of the node. All of these components are independent from one another\footnote{Another way to see this is to remember that the labelling of the nodes can be rearranged so that $\Ht$ becomes a block diagonal matrix, with each block corresponding to a connected component. When removing the rows and columns associated to a node, the block to which it belongs gets divided in $k_i$ different ones (a further permutation might be required to make this explicit), and when inverting $\Ht$ each of these blocks can be inverted independently, thus yielding independent $\{\G_{jk}^{(j)}\}$. }, and hence removing one of $j$'s neighbours after creating the cavity does not affect the others. 
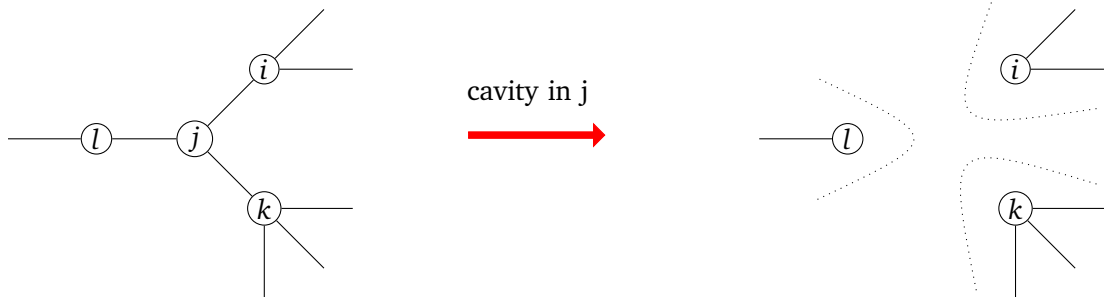
\begin{figure}[t]
    \centering
    \begin{tikzpicture}[node distance={13mm}, main/.style = {draw, circle, inner sep=1pt}] 
\node[main]  (j) at (0,1) {$j$}; 
\node[main] [above right of=j] (i) {$i$};
\node[main] [below right of=j] (k) {$k$};
\node[main] [left of=j] (l) {$l$};
\node[above right of=i] (ip) {};
\node[right of=i] (is) {};
\node[below of=k] (kp) {};
\node[right of=k] (ks) {};
\node[below right of=k] (kt) {};
\node[left of=l] (lp) {};
\draw (j) -- (i);
\draw (j) -- (k);
\draw (j) -- (l);
\draw (i) -- (ip);
\draw (i) -- (is);
\draw (k) -- (kp);
\draw (k) -- (ks);
\draw (k) -- (kt);
\draw (l) -- (lp);
\node[single arrow, draw=red, fill=red, 
       minimum width = 2pt, single arrow head extend=3pt,
      minimum height=18mm,inner sep =1pt,above] at (4.5,1) (arr) {};
\node[text width=18mm] at (4.5,1.6) {cavity in j};
\end{tikzpicture} \hfill%
\begin{tikzpicture}[node distance={13mm}, main/.style = {draw, circle, inner sep=1pt}]
\draw[thin,dotted] (0.6,2.8) .. controls (0,1.1) .. (2,1.4);
\draw[thin,dotted] (0.4,-1) .. controls (0,1) .. (2,0.4);
\draw[thin,dotted] (-1.7,0.2) .. controls (0,1) .. (-1.7,1.8);
\node[]  (j) at (0,1) {}; 
\node[main] [above right of=j] (i) {$i$};
\node[main] [below right of=j] (k) {$k$};
\node[main] [left of=j] (l) {$l$};
\node[above right of=i] (ip) {};
\node[right of=i] (is) {};
\node[below of=k] (kp) {};
\node[right of=k] (ks) {};
\node[below right of=k] (kt) {};
\node[left of=l] (lp) {};
\draw (i) -- (ip);
\draw (i) -- (is);
\draw (k) -- (kp);
\draw (k) -- (ks);
\draw (k) -- (kt);
\draw (l) -- (lp);
\end{tikzpicture}
    \caption{Explicative diagram showing the idea behind \eqref{SI-BreakingHier}.}
    \label{SI-fig:breakinghier}
\end{figure}
\noindent
By applying \eqref{SI-BreakingHier} we finally get the closed-form equations we were looking for,
\begin{equation}\label{SI-eq:finalgcav_supplemental}
    \G^{(l)}_{jj}=-\bigg(\z-\M_{jj}+ \sum_{k\in\partial_j	\setminus\{l\}}\M_{jk}\G^{(j)}_{kk}\M_{kj}\bigg)^{-1},
\end{equation}
from which we can obtain the cavity blocks $\{\G^{(l)}_{jj}\}$ and then the ones defined on the original graph, $\G_{jj}$. Note that the problem of inverting the $2N\times 2N$ auxiliary matrix $\ub{H}$ has been now reduced to the inversion of $N$ blocks of dimensions $2\times2$, which is significantly easier.

These equations were derived under the assumption of having a tree graph, whereas \ER \,graphs are only locally-treelike, hence by using Eq.~\eqref{SI-eq:finalgcav_supplemental} for our model we are actually making an approximation. However, in the infinite size limit it is safe to assume that cycles of diverging length introduce negligible correlations to the cavity blocks $\{\G_{jk}^{(j)}\}$, and that the contributions brought by cycles of finite length, the number of which stays finite when $N\to\infty$, can be ignored. Under these assumptions, Eq.~\eqref{SI-eq:finalgcav_supplemental} become exact for locally-treelike graphs in the infinite size limit, therefore validating their use for our analysis.

\subsection{The trivial solution to the cavity equations and the boundary of the support of the spectrum}
We now discuss a particular solution to the equations we have derived before, called the \emph{trivial solution} \cite{metz2012spectra, metz2019spectral}, which is obtained by assuming a diagonal form for the $\G_{jj}$-s and the $\G^{(l)}_{jj}$-s, \ie
\begin{equation}
    \G^0_{jj}=
    \begin{pmatrix}
        g_j & 0\\
        0 & \overline{g}_j
    \end{pmatrix}, \quad \G^{0,(l)}_{jj}=
    \begin{pmatrix}
        g^{(l)}_j & 0\\
        0 & \overline{g}^{(l)}_j
    \end{pmatrix}.\label{SI-eq:trivialG-supplemental}
\end{equation}
where the superscript $^0$ is used to distinguish the trivial solution from the other possible ones. These expressions are the simplest solutions that can be obtained when considering the case $\eta=0$ in the auxiliary matrix defined in Eq.~\eqref{SI-auxiliary}, thus making all of the elements belonging to its off-diagonal $N\times N$ blocks equal to 0. We can now see that our choice of the auxiliary matrix leads to this simple, diagonal, form of the trivial solution.  We stress that this form differs from the one typically found in literature, which is obtained by putting the diagonal elements to 0 rather than the off-diagonal ones: this is due to the fact that it is often preferred a form of $\ub{H}$ which is Hermitian for $\eta=0$, in which  $(\textbf{M}-z\textbf{1}_N)$ and $ (\textbf{M}-z\textbf{1}_N)^\dag$ are placed in the off-diagonal $N\times N$ blocks, as opposed to our case.

If we stick to the case $\eta=0$ and plug this ansatz into the cavity equations, we immediately obtain the ones for the $g_j-$s and the $g^{(l)}_j-$s, namely
\begin{gather}
    g_{j}=\frac{-1}{z-[\textbf{M}]_{jj}+ \displaystyle{\sum_{k\in\partial_j}}[\textbf{M}]_{jk}g^{(j)}_{k}[\textbf{M}]_{kj}}\label{SI-trivial},\\
    g^{(l)}_{j}=\frac{-1}{z-[\textbf{M}]_{jj}+ \displaystyle\sum_{k\in\partial_j	\setminus\{l\}}[\textbf{M}]_{jk}g^{(j)}_{k}[\textbf{M}]_{kj}}\label{SI-eq:trivialcav_supplemental}.
\end{gather}
These simple expressions do not hold when $\eta\neq0$, but they still do, up to an error linear in $\eta$, when we send $\eta\to0$
As it can be seen by applying Schur's formula to Eq.~\eqref{SI-Resolvent}, these equations are the same that describe the diagonal elements of the resolvents\footnote{We denote by $\textbf{M}^{(l)}$ the matrix built on the cavity graph $\mathscr{G}^{(l)}$.} $\textbf{G}_{\textbf{M}}(z)$ and $\textbf{G}_{\textbf{M}^{(l)}}(z)$, and, just like them, they are well-defined only outside of the spectra of $\textbf{M}$ and $\textbf{M}^{(l)}$. Moreover, if we put $z=x+i\tilde{\eta}$ with $\tilde{\eta}\in \mathbb{R}$ as in Eq.~\eqref{SI-regul}, we immediately get the cavity equations for the regularised resolvent of symmetric matrices, which have been derived in Ref.~\cite{rogers2008cavity}. As discussed in Refs.~\cite{metz2019spectral} and \cite{neri2016eigenvalue}, in the infinite size limit the trivial solution is only valid outside of a region of the complex plane $\tilde{\mathscr{S}}$ which in most cases coincides with the spectral bulk. Inside this domain, the solution is no longer the correct one, and this translates to its instability with respect to small off-diagonal perturbations, as the ones produced for example by a nonzero $\eta$. A linear stability analysis can hence be used to determine whether a point in $\mathbb{C}$ belongs to the spectral bulk or not, and this idea lies at the basis of the numerical studies presented in the main text.

In order to find a criterion to assess the stability of the trivial solution, we start by adding two small off-diagonal perturbations to each $\G^{0,(l)}_{jj}$, thus obtaining
\begin{equation}
    \G^{p,(l)}_{jj}=
    \begin{pmatrix}
        g^{(l)}_j & \delta_j^{(l)}\\
        \epsilon_j^{(l)} & \overline{g_j}^{(l)}
    \end{pmatrix},
\end{equation}
where $\delta_j^{(l)}$ and $\epsilon_j^{(l)}$ are small real numbers, and the superscript p is used to identify the perturbed trivial solution. By using this expression, the cavity equations \eqref{SI-eq:finalgcav_supplemental} now read as

\begin{adjustbox}{max width=0.905\textwidth}
\parbox{\linewidth}{%
\begin{gather}
    -\bigg(\begin{pmatrix}
        z-[\textbf{M}]_{jj} & 0 \\
        0 & \overline{z}-\overline{[\textbf{M}]_{jj}}
    \end{pmatrix}+\sum_{k\in \partial_j\setminus\{l\}}
    \begin{pmatrix}
       [\textbf{M}]_{jk} &0\\
       0 &\overline{[\textbf{M}]_{kj}}
   \end{pmatrix}
   \begin{pmatrix}
        g_k^{(j)} & \delta_{k}^{(j)} \\
        \epsilon_{k}^{(j)} & \overline{g_k }^{(j)}
    \end{pmatrix}
   \begin{pmatrix}
       [\textbf{M}]_{kj} &0\\
       0 &\overline{[\textbf{M}]_{jk}}
   \end{pmatrix}
    \bigg)^{-1}=\displaybreak[0]\nonumber\end{gather} 
}
\end{adjustbox}
\begin{gather}
    =-\begin{pmatrix}
        z-[\textbf{M}]_{jj}+\displaystyle\sum_{k\in \partial_j\setminus\{l\}}[\textbf{M}]_{jk} g_k^{(j)}[\textbf{M}]_{kj} & \displaystyle\sum_{k\in \partial_j\setminus\{l\}}\delta_{k}^{(j)} |[\textbf{M}]_{jk}|^2\\\\
        \displaystyle\sum_{k\in \partial_j\setminus\{l\}}\epsilon_{k}^{(j)} |[\textbf{M}]_{kj}|^2 & \overline{z}-\overline{[\textbf{M}]_{jj}}+\displaystyle\sum_{k\in \partial_j\setminus\{l\}}\overline{[\textbf{M}]_{jk}} \overline{g}_k^{(j)}\overline{[\textbf{M}]_{kj} }
    \end{pmatrix}^{-1}=\nonumber\\
    \frac{-1}{D_1-D_2}\begin{pmatrix}
         \overline{z}-\overline{[\textbf{M}]_{jj}}+\displaystyle\sum_{k\in \partial_j\setminus\{l\}}\overline{[\textbf{M}]_{jk}} \overline{g}_k^{(j)}\overline{[\textbf{M}]_{kj}}& -\displaystyle\sum_{k\in \partial_j\setminus\{l\}}\delta_{k}^{(j)} |[\textbf{M}]_{jk}|^2\\\\
        -\displaystyle\sum_{k\in \partial_j\setminus\{l\}}\epsilon_{k}^{(j)} |[\textbf{M}]_{kj}|^2 &  z-[\textbf{M}]_{jj}+\displaystyle\sum_{k\in \partial_j\setminus\{l\}}[\textbf{M}]_{jk} g_k^{(j)}[\textbf{M}]_{kj}
    \end{pmatrix},
\end{gather}
where we have used the standard formula for the inverse of a $2\times2$ matrix and  defined the following quantities:
\begin{gather}
    D_1=\bigg|z-[\textbf{M}]_{jj}+\displaystyle\sum_{k\in \partial_j\setminus\{l\}}[\textbf{M}]_{jk} g_k^{(j)}[\textbf{M}]_{kj}\bigg|^2=\bigg|g_j^{(l)}\bigg|^{-2},\\
    D_2=\Bigg(\displaystyle\sum_{k\in \partial_j\setminus\{l\}}\epsilon_{k}^{(j)}|[\textbf{M}]_{kj}|^2\Bigg)\Bigg(\displaystyle\sum_{k\in \partial_j\setminus\{l\}}\delta_{k}^{(j)}|[\textbf{M}]_{jk}|^2\Bigg)=O(\epsilon\delta) \, .
\end{gather}
Notice that in the last expression we have omitted the indices from $O(\epsilon\delta)$ in order to simplify the notation. From these definitions, it is straightforward to see that
\begin{equation}
    \frac{1}{D_1-D_2}=\bigg|g_j^{(l)}\bigg|^{2}+O(\epsilon\delta) \, ,
\end{equation}
and hence it is easy to derive equations for the off-diagonal terms up to the linear order in both $\epsilon$ and $\delta$, namely
\begin{gather}
\delta_{j}^{(l)}=\bigg|g_j^{(l)}\bigg|^{2}\displaystyle\sum_{k\in \partial_j\setminus\{l\}}\delta_{k}^{(j)} |[\textbf{M}]_{jk}|^2, \label{SI-eq:epsilon_supplemental}\\
\epsilon_{j}^{(l)}=\bigg|g_j^{(l)}\bigg|^{2}\displaystyle\sum_{k\in \partial_j\setminus\{l\}}\epsilon_{k}^{(j)} |[\textbf{M}]_{kj}|^2,\label{SI-eq:delta_supplemental}
\end{gather}
where the $g_j^{(l)}$-s obey Eq.~\eqref{SI-eq:trivialcav_supplemental}. When these quantities go to zero, the trivial solution is stable, and therefore we deduce that the point $z$ we are considering is located outside the bulk. On the other hand, when the quantities in Eqs.~\ref{SI-eq:epsilon_supplemental} and \ref{SI-eq:delta_supplemental} diverge, we conclude that $z$ belongs to the bulk.